\documentclass[twocolumn,showpacs,prx,eqsecnum,citeautoscript,amsmath,amssymb,floatfix,superscriptaddress]{revtex4-1}
\usepackage{bm,xcolor,amsmath,amssymb,mathrsfs,latexsym,graphicx,psfrag,float,dsfont}
\usepackage{dcolumn}
\usepackage{hyperref}

\usepackage{leftidx}

\usepackage{tabularx}

\usepackage[english]{babel}
\usepackage[utf8]{inputenc}
\usepackage{bbold}

\usepackage{tikz}
\usetikzlibrary{arrows,decorations.pathmorphing,backgrounds,positioning,fit,matrix}

\usepackage[normal]{subfigure}		

\newcommand{\lan}{\left\langle}
\newcommand{\ran}{\right\rangle}

\newcommand{\matb}{\left(\begin{array}}
\newcommand{\mate}{\end{array}\right)}
\newcommand{\sysb}{\left\{\begin{array}}
\newcommand{\syse}{\end{array}\right.}
\newcommand{\xif}{\boldsymbol{\xi}}

\newcommand{\ket}[1]{\left| #1 \ran}
\newcommand{\bra}[1]{\lan #1 \right|}

\newcolumntype{L}[1]{>{\raggedright\arraybackslash}p{#1}} 
\newcolumntype{C}[1]{>{\centering\arraybackslash}p{#1}} 
\newcolumntype{R}[1]{>{\raggedleft\arraybackslash}p{#1}} 

\newcommand{\be}{\begin{equation}}
\newcommand{\ee}{\end{equation}}

\newcommand{\eq}[2]{\begin{align}\label{#1} #2 \end{align}}

\newcommand{\fdif}[2]{\ensuremath{\frac{\delta #1}{\delta #2}}}

\begin{document}

\title{Phenomenology of a First Order Dark State Phase Transition}

\author{Dietrich Roscher}
\affiliation{Institut f\"ur Theoretische Physik, Universit\"at zu K\"oln, D-50937 Cologne, Germany}
\affiliation{Department of Physics, Simon Fraser University, Burnaby, British Columbia, Canada V5A 1S6}
\author{Sebastian Diehl}
\affiliation{Institut f\"ur Theoretische Physik, Universit\"at zu K\"oln, D-50937 Cologne, Germany}
\author{Michael Buchhold}
\affiliation{Department of Physics and Institute for Quantum Information and Matter, California Institute of Technology, Pasadena, CA 91125, USA}

\date{\today}

\begin{abstract}
Dark states are stationary states of a dissipative, Lindblad-type time evolution with zero von Neumann entropy, therefore representing examples of pure, steady quantum states. Non-equilibrium dynamics featuring a dark state recently gained a lot of attraction since their implementation in the context of driven-open quantum systems represents a viable possibility to engineer unique, pure states. In this work, we analyze a driven many-body spin system, which undergoes a transition from a dark steady state to a mixed steady state as a function of the driving strength. This transition connects a zero entropy (dark) state with a finite entropy (mixed) state and thus goes beyond the realm of equilibrium statistical mechanics and becomes of genuine nonequilibrium character. We analyze the relevant long wavelength fluctuations driving this transition in a regime where the system performs a discontinuous jump from a dark to a mixed state by means of the renormalization group. This allows us to approach the nonequilibrium dark state transition and identify similarities and clear differences to common, equilibrium phase transitions, and to establish the phenomenology for a first order dark state phase transition.
\end{abstract}

\maketitle
\section{Introduction}
Pure states play an important role in the understanding of phases of matter and the dynamics close to a phase transition for a large number of many-body systems \cite{QPT}. While many physical observations can be understood from the ground state properties of a given Hamiltonian, realistic systems are never perfectly isolated and at the same time at zero entropy. Observables in generic systems are therefore subject to statistical fluctuations, resulting from the statistical mixture of several pure states, which e.g. overlay the inherent quantum fluctuations of a single pure state \cite{QPT, QPT1}. Important examples for which this becomes apparent are found in the long-time and long-wavelength dynamics of interacting quantum systems, e.g. in the emergence of classical criticality \cite{QPT2, QPT3} and classical hydrodynamics at finite temperatures \cite{HydroMit, Lux14}. 

The investigation of pure many-body states has gained fresh impetus lately by realizing that instead of cooling a system towards its exact ground state by slowly reducing its temperature, it is possible to push it into a pure state by adding external drive and dissipation to its dynamics \cite{DiehlDark1, Dark2, DiehlDark3}. Balancing drive and loss channels for a given system in the right way, it was shown that indeed it may relax towards a so-called dark state, i.e. a pure state which is no longer susceptible to drive and dissipation \cite{Dark1, DiehlDark2, Dark3, Dark4}. Within this framework one can engineer both rather "classical" product states such as ferromagnets, Ne\' el states and non-interacting Bose-Einstein condensates as well as genuine quantum states, including topologically non-trivial ones. 

While the initial focus has been set on the engineering of particular dark states \cite{DiehlDark1, DiehlDark3}, it is a natural question to ask for the nature of the many-body dynamics in the vicinity of a dark state transition. This requires a setup, which features the existence of dark states for an extended parameter regime. We consider driven open Rydberg systems for which, by explicitly varying the drive strength versus the dissipation in the system, the asymptotic stationary state undergoes a transition from a mixed state, which has a non-vanishing von Neumann entropy $S>0$, to a pure state, which has zero entropy $S=0$. 

This brings phase transitions between a dark state and a mixed state phase into contact with nonequilibrium statistical physics, where classical steady states with zero entropy are known as absorbing states \cite{DP_Janssen, Janssen2005, DP_Hinrichsen, Marcuzzi2015, Marcuzzi2016, NoiseFRGI}. Continuous phase transitions into absorbing states have been extensively studied theoretically since they are believed to model a diverse set of dynamical processes such as disease spreading, the evolution of forest fires, percolation processes and population dynamics \cite{Janssen1999, NEQ_PT1,Hinrichsen_exp}. Key feature to these models is a phase characterized by a fluctuationless, zero-entropy state, separated from a fluctuating phase of non-zero entropy. This denies the presence of a temperature scale and makes the phase transition into absorbing states a genuine nonequilibrium transition \cite{Tauber1, Tauber-book}.

Experimental realizations of absorbing state phase transitions using "quantum ingredients" have been proposed in the framework of driven Rydberg systems very recently for both incoherent and coherent drive \cite{Marcuzzi2015, Marcuzzi2016, Valado2015}. These absorbing states have, however, not been explicitly identified as dark states as a large part of the work was carried out in the classical limit \cite{NoiseFRGI}. We build up on these proposals and study the dynamics of coherently driven Rydberg ensembles, for which the notion of a dark state phase transition is most appealing. 

Mean-field theory and small-scale numerics separate the steady states into a dark state for small external drive and a mixed state for large external drive and identify a region of bistability between both states at intermediate drive strengths, which suggests that both phases are separated by a first order phase transition in the thermodynamic limit of infinite system size \cite{Marcuzzi2016, NoiseFRGI}.

Understanding the dynamics at first order phase transitions represents a general challenge for theoretical physics \cite{BinderRev, Langer1}. The interest in dynamics in a bistable regime and first order phase transitions has only recently started growing again, initialized by the observation that many driven dissipative many-body systems can feature bistable dynamics in regimes of strong drive and dissipation \cite{Angerer2017,NoiseFRGI,Rodriguez2017,Hruby2017}. These systems have, however, in common that in the bistable regime, the noise level approaches a finite, constant value for both metastable states, which corresponds to a dissipative phase transition at fixed, finite von Neumann entropy and thus to an effective, dissipative equilibrium. 

The dynamics in a bistability region between a dark state and a mixed steady state is very special in several respects. Phenomenologically, the transition from one thermodynamic phase to another via a first order transition is understood in terms of the nucleation of small droplets and their subsequent growth to a macroscopic size \cite{BinderRev, Langer1, Langer2}. In the dark state framework this implies that large regions of zero entropy as well as non-vanishing entropy coexist close to a first order dark state transition, a genuinely nonequilibrium circumstance. Exactly at the transition, however, both the zero and the finite entropy phase coexist, which translates to the coexistence of pure as well as mixed steady states in a density matrix picture. Even the classical counterparts, first order phase transitions into absorbing states, are hardly understood on a qualitative level and believed to be very rare in higher dimensions and even impossible in one dimension \cite{1Ddisc, Grassberger, 1dperc, Dickman1991}. 

The intention of this work is to establish the phenomenology for a first order dark state transition in dependence of spatial dimensionality. We also introduce a functional renormalization group \cite{Wetterich1993,Berges2002,Canet2004} based approach, tailored to investigate long-wavelength dynamics of large order parameter fluctuations in a non-equilibrium setting to the currently developing theoretical toolbox for driven dissipative dynamics \cite{Fazio2016, Ciuti2015, Biella2018, Lange2017}. This approach works in both low and high dimensions and allows us to resolve the rather complex but also very rich and fascinating dynamics close to a first order phase transition as well as a fluctuation induced second order phase transition, which we highlight on the following pages. 

The dark state can be understood as a simple ferromagnetic product state and is thus particularly well suited for the investigation of the phase transition. As we will show, this allows us to construct an order parameter field that is free of any fluctuations, classical and quantum, when the system is in the dark state. Thus the defining property of a dark state, i.e. the absence of statistical fluctuations, is not masked by remaining quantum fluctuations of the order parameter.
The quantum nature of the system nevertheless comes crucially into play at two different stages. First, non-vanishing coherences in the local spin density matrix are necessary to make a discontinuous transition in this system at all possible \cite{NoiseFRGI} and second, the absence of any statistical fluctuation scale requires a pure state. 

In order to analyze the steady state properties and the dynamics in the bistable regime, we identify the local density of atoms in the Rydberg state as the order parameter. Its expectation value {\bf and} fluctuations vanishes exactly for a dark state. Its dynamics will be described in terms of an effective, nonequilibrium field theory, which we solve via a semi-analytical approach based on the functional Renormalization Group (fRG). Relating the obtained results with the phenomenology of thermal first order transitions allows us to extract a number of physical insights regarding the nonequilibrium setting. We benchmark these findings with numerically exact results in one spatial dimension, thereby establishing the method as a general tool to study first order nonequilibrium phase transitions.
\begin{figure}
\centering
\includegraphics[width=0.5\textwidth]{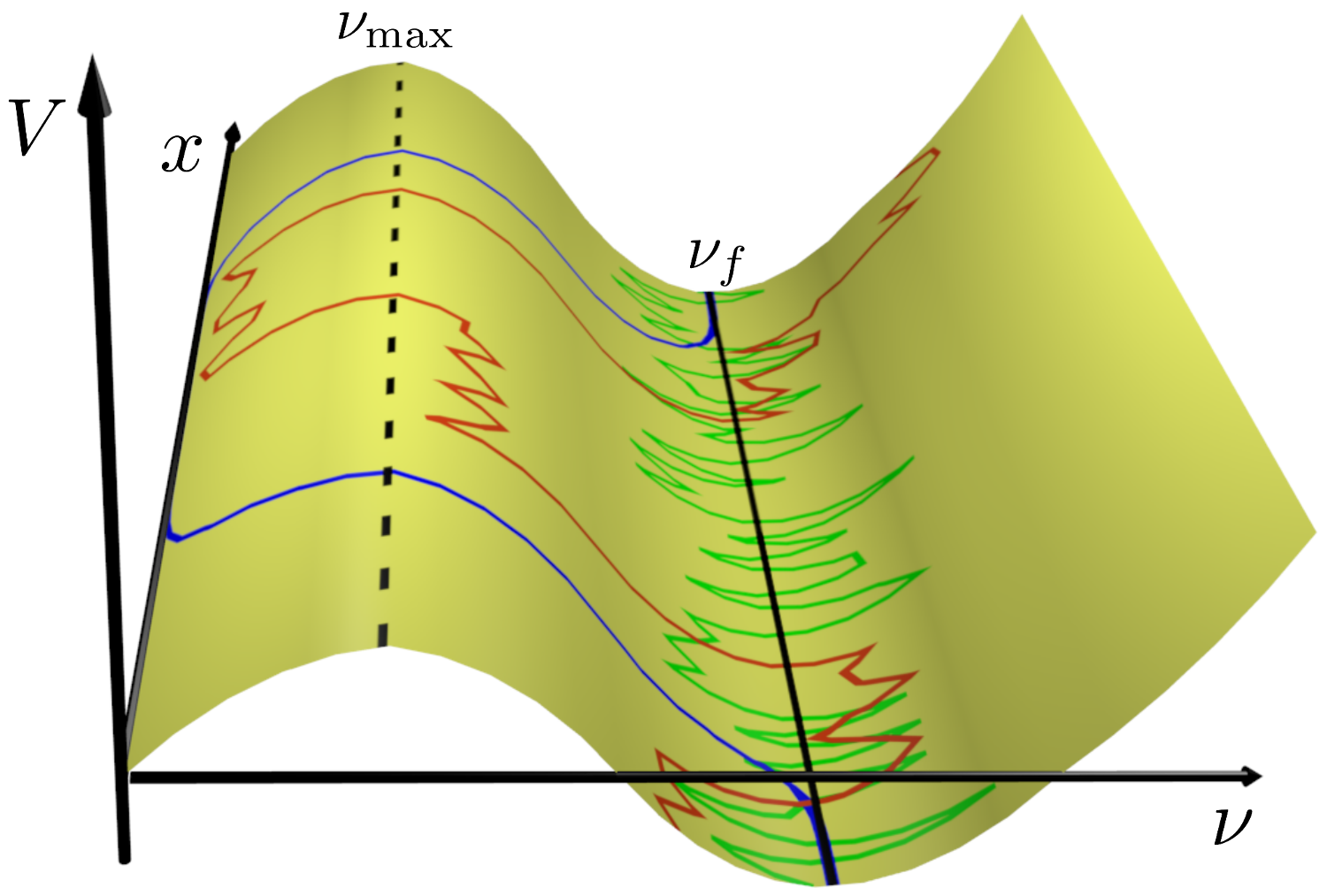}
\caption{Portrait of prototypical inhomogeneous configurations of the coarse grained excitation density $\nu_x$ in one spatial dimension (labeled with $x$) that fluctuate about the nontrivial minimum at $\nu_x=\nu_f$. At each point in space $x$ the field has a value $\nu_x$ and experiences the potential $V(\nu_x)$, which is maximal at $V(\nu_{\rm{max}})$. Perturbatively small amplitude oscillations (green) do not probe global properties of $V$. In contrast, droplet-like saddle point solutions (blue) or strongly fluctuating configurations (red) are crucial to understand the dominant behavior at the first order and the second order phase transition, as discussed below.}\label{figPot}
\end{figure}

\subsection{Summary of Results}
Starting from the quantum master equation for the driven dissipative Rydberg setup \cite{Marcuzzi2015, Marcuzzi2016, Weimer2010}, we introduce the density of atoms in the Rydberg state $\nu_{X}\ge0$, $X=(x,t)$ as the order parameter field for the dark state transition in $d$ spatial dimensions. Its dynamics follows the fundamental Langevin equation
\eq{Langevin}{
\partial_t\nu_{X}=D\nabla^2\nu_{X}-V'(\nu_X)+\xi_{X}
}
of diffusive motion in a potential $V$ subject to a multiplicative, Gaussian noise $\xi_X$. It has zero mean $\langle \xi_X\rangle=0$ and variance $\langle \xi_X\xi_Y\rangle = \delta(X-Y)\nu_X$. The double-well potential $V$ is illustrated in Fig.~\ref{figPot}. It has two minima at $\nu_X=0$ and $\nu_X=\nu_f>0$, which are separated by a well that is maximal at $\nu_X=\nu_\text{max}$.

Considering the general importance of fluctuations on all length scales for a first order phase transition, we establish a fRG approach suitable for non-equilibrium phase transitions. This approach evolves both the potential {\bf and} the noise kernel in a non-trivial way, which turns out to be of major importance for the resolution of droplet formation close to the first order phase transition, and allows us to develop a clear picture of the critical dynamics.

We demonstrate that both a dark state phase as well as a mixed state phase can be stabilized in the presence of spatial fluctuations and nonequilibrium noise, and we quantitatively determine the location of the phase boundary as a function of dimensionality and the strength of the noise level. 
The nature of the phase transition is sensitive to the dimensionality of the system and the height of the potential barrier. In general one can distinguish two cases, a true first order phase transition, accompanied with a discontinuous jump of the order parameter on the one hand and a fluctuation induced, second order transition, for which spatial fluctuations have softened the evolution of the order parameter towards a continuous evolution at the critical point on the other hand.

The second order phase transition is especially pronounced in one dimension, where spatial fluctuations are so strongly enhanced that they can overcome the potential barrier on intermediate distances and lead to a continuous evolution of $\nu_X$ at the phase transition in the entire parameter regime. This can be directly traced with the fRG flow of the potential, where fluctuations on intermediate wavelengths render the initial double well convex and establish a single, global minimum for $V$. The dark and mixed state are no longer separated by a potential barrier at these scales and the long-wavelength modes fluctuate randomly between the two solutions (red configuration in Fig.~\ref{figPot}). Scaling dynamics, however, persists onto the largest length scales and the single potential minimum evolves towards zero under the influence of long distance modes, thus undergoing a continuous transition. For this and other exemplary situations, we provide animated data of the RG flow~\cite{SM}. We find that the manifold and rather complex results are better conveyed in this way than with individual, static plots only.

In higher dimensions $d>1$, the potential barrier persists onto the largest scales and vanishes only in the limit of long-wavelengths, establishing a single minimum that jumps when passing the transition. This hints towards a first order transition induced by the nucleation and growth of droplets, i.e. meta-stable field configurations, which interpolate smoothly between the two phases (blue configuration in Fig.~\ref{figPot}) and thus require the presence of a potential barrier on all but the lowest scales \cite{Langer1, Langer2}. 

This picture of the phase transitions is confirmed by the RG evolution of the noise kernel. At the first order transition, both the onset of the potential evolution as well as a sudden deformation of the noise kernel are observed at a very sharp, intermediate length scale. The deformed noise establishes a bimodal structure with one maximum on each side of the potential barrier, indicating strongly pronounced fluctuations of the order parameter between the potential minima, see Fig.~\ref{TransNoise}. Both observations are in accordance with the picture of a first order phase transition driven by the formation of droplets of well defined extend \cite{Langer1, Langer2,BinderRev}. Droplets form above a sharp, intermediate length scale and perform sudden jumps from one phase to the other, i.e. between the two potential minima, leading to an increased fluctuation rate at these values. In contrast, in $d=1$, neither a sharp length scale for the onset of the evolution nor a pronounced bimodal structure of the noise kernel can be identified, pointing out the crucial difference between fluctuations leading to first or second order phase transitions.

At the first order transition point, we observe the formation of a completely flat, coarse grained potential {\bf and} noise kernel $\chi=V=0$ for an extended region of field configurations $0\le \nu_X\le \nu_f$. In the Langevin framework, this is identified as an extensive number of stationary, noiseless field configurations and the emergence of the coexistence of the dark state and the mixed state, i.e. of a zero entropy and finite entropy phase at this point. In a master equation framework this remarkable observation hints towards an extensive number of dark states at the coexistence point, each corresponding to a specific droplet configuration. This is contrasted by the common observation of {\bf two} degenerate steady states at a coexistence point with constant but finite noise level \cite{Kessler2012, Casteels2017}. 

\section{Dark State Spin Model}
\label{sec:ModelBuild}
The opportunity to precisely manipulate entire ensembles of atomic spins with quantum optical tools, such as coherent drive and pump lasers \cite{Ryd-lattice1,Ryd-lattice2,Endres2017, Hofmann2013, SiebererReview, Carusotto13}, as well as with dissipative jump operators via reservoir engineering \cite{Poyatos96, Bose99,Krauter2011, Schindler2011}, pushed forward the search for robust, many-body dark states. The latter are pure states $\rho_{\text{DS}}=|\text{DS}\rangle\langle\text{DS}|$, which are exact, stationary zero modes $\mathcal{L}\rho_{\text{DS}}=0$ of the quantum master equation, $\partial_t\rho=\mathcal{L}\rho$. Once a dark state is reached during the time evolution, the system will remain forever in this particular state and the dynamics has terminated. In this sense, $\rho_{\text{DS}}$ represents the quantum mechanical analogue of a classical absorbing state, for which both the deterministic time evolution as well as reservoir induced fluctuations vanish. The von Neumann entropy $S(\rho)=-\text{tr}\left(\rho\log\rho\right)$ of a dark state is always zero $S(\rho_{\text{DS}})=-\log 1=0$.

Similar to a fluctuation-less absorbing state, dark states are very clean representatives of steady states, which feature the complete absence of statistical (but not necessarily quantum) fluctuations. The mere existence of a dark state $\rho_{\text{DS}}$ in a given model does, however, not imply that $\rho_{\text{DS}}$ is the only attractor of the dynamics for this model \cite{Marcuzzi2016, NoiseFRGI}. Under certain conditions, which we will elaborate on in more detail below, it may be that in addition to a unique dark state, there exists another, mixed steady state $\rho_{\text{SS}}=\rho_{MS}=\sum _\alpha p_{\alpha}|\psi_\alpha\rangle\langle\psi_\alpha|\neq\rho_{\text{DS}}$, such that each initial state $\rho(t=0)\neq\rho_{\text{DS}}$ evolves towards $\rho_{\text{SS}}$ asymptotically, $\rho(t)\overset{t\rightarrow\infty}{\rightarrow}\rho_{\text{SS}}$. This mixed state has per definition non-vanishing von Neumann entropy $S(\rho_{MS})=-\sum_\alpha p_\alpha\log p_\alpha>0$. 

\subsection{Quantum master equation}
A minimal model, which displays a transition from a pure steady state towards a fluctuating steady state has been proposed in Refs.~\cite{Marcuzzi2016, NoiseFRGI}. It describes the time evolution of a spin ensemble on a $d$-dimensional square lattice subject to coherent and dissipative drive. Here, each spin-$\frac{1}{2}$ degree of freedom represents the ground $|g\rangle\equiv|\downarrow\rangle$ or the excited $|e\rangle\equiv|\uparrow\rangle$ state of a Rydberg atom, as used in a number of recent experiments \cite{Endres2017, Amthor2010, Malossi14, Helmrich2016, Beguin2013}. The key to achieve true nonequilibrium dynamics is the exploitation of the so-called anti-blockade mechanism in driven Rydberg ensembles, which, by resonant driving, facilitates the excitation of atoms in the vicinity of other Rydberg-excited atoms. At the same time it suppresses the excitation rate of atoms in the absence of nearby Rydberg states \cite{Lee2012, Ryd-QI, Urvoy14}. A detailed discussion of the experimental implementation can be found in Refs.~\cite{Amthor2010,Valado2015,Gutierrez2016}.

In this work, we focus on the case of a pure coherent drive, which, on the mean-field level, features a first order phase transition from a pure state containing no spin excitations to a fluctuating steady state with a non-vanishing spin excitation density \cite{Marcuzzi2016, NoiseFRGI}. This regime has not yet been explored beyond the mean-field level. It is, however, within reach of current experiments with driven Rydberg ensembles, which can potentially probe the nature of the nonequilibrium phase transition \cite{Gutierrez2016, Helmrich2016}, and thus its theoretical understanding is pressing. 

We consider an infinite, $d$-dimensional square lattice with lattice spacing $a$. Each site, labeled with the index $l$, hosts a single spin, represented by the basis states ${|\downarrow\rangle_l, |\uparrow\rangle_l}$ and a set of Pauli-operators $\sigma^{x,y,z}_l$. The spins are coherently driven by a laser with Rabi frequency $\Omega$ and detuning $\Delta=-V_{nn}$, where $V_{nn}$ is the nearest neighbor Rydberg repulsion. The Rydberg repulsion is by far the strongest scale and thus transitions between the ground and the Rydberg state are only resonant in the vicinity of another excited atom, while they are far off resonance in the absence of nearby Rydberg excitations. Projecting onto the resonant transitions, the atomic level splitting, Rydberg repulsion and laser drive are summarized in the Hamiltonian \cite{Marcuzzi2016, NoiseFRGI}  
\eq{eq5}{
H=\Omega\sum_l \Pi_l\sigma^x_l.
}
Here, $\Pi_l$ is the projector~\footnote{In contrast to the definition in Refs.~\cite{Marcuzzi2016,NoiseFRGI}, $\Pi_l$ as defined in the present work is a true projector. While this is more accurately modeling the quantum contact process, it does, however, not change the long wavelength physics compared to the previously used definitions.} onto the states with at least one up-spin in the neighborhood of site $l$
\eq{eq3}{
\Pi_l=\mathds{1}-\prod_{m\text{ nn }l}(\mathds{1}-n_m), \text{ with } n_m=\frac{\sigma^z_m+1}{2}.
} 
For a state $|\psi\rangle$ one finds $\Pi_l|\psi\rangle=0$ if $n_m|\psi\rangle=0$ for all nearest neighbors ("nn") $m$ of $l$ and $\Pi_l|\psi\rangle=|\psi\rangle$ else. 

The highly excited Rydberg state can decay back into the ground state by spontaneously emitting a photon, this incoherent process is well described by a Markovian master equation with a dissipator proportional to the loss rate $\gamma$.
The master equation for the density matrix is
\eq{eq4}{
\partial_t\rho=\mathcal{L}\rho=i[\rho,H]+\gamma\sum_l\left(\sigma^-_l\rho \sigma^+_l-\frac{1}{2}\{n_l,\rho\}\right),
}
with the common spin ladder operators $\sigma^{\pm}_l$ and the Liouvillian superoperator $\mathcal{L}$ summarizing the RHS.

The quantum master equation features one dark state independently of the choice of the coupling constants $\{\Omega, \gamma\}$. This is the ferromagnetic spin-down state 
\eq{eq10}{
\rho_{\text{DS}}=|\psi_\downarrow\rangle\langle\psi_\downarrow|, \text{ with } |\psi_\downarrow\rangle=\prod_l|\downarrow\rangle_l,
}
and it is easy to see that $\mathcal{L}\rho_{\text{DS}}=0$. While this state, however, is present throughout the entire parameter regime, we will show that there exists a large domain $\mathcal{A}$ in parameter space, for which $\rho_{\text{DS}}$ is not attracting the dynamics and becomes fully repulsive in the thermodynamic limit. In other words, for $(\Omega,\gamma)\in\mathcal{A}$ any initial state $\rho_0$ that is different from $\rho_{\text{DS}}$ will evolve towards a steady state $\rho_{\text{SS}}\neq\rho_{\text{DS}}$. Within this domain, the dark state represents an isolated state, which can neither be reached nor left dynamically. Crossing the boundary of $\mathcal{A}$ is connected to a phase transition from $\rho_{\text{SS}}\neq\rho_{\text{DS}}$ towards $\rho_{\text{SS}}=\rho_{\text{DS}}$.

\subsection{Rydberg density Heisenberg-Langevin equation}
The dark state $\rho_{\text{DS}}$ is an eigenstate of the local density
\eq{eq11}{
{n}_l\rho_{\text{DS}}=0,
} 
which corresponds to a vanishing density of Rydberg excitations. Since for integer $m$ one finds $n_l^m=n_l$, this implies that all moments of the density are zero in the dark state. It is a fluctuationless state with respect to $n_l$. It is evident that for any mixed state, $\langle n_l\rangle>0$ for some $l$ and thus $\langle n_l\rangle$ serves well as an order parameter for the dark state transition. In order to investigate its dynamics, we derive now the corresponding field theory. 
We take a different route than Refs.~\cite{Marcuzzi2016, NoiseFRGI} and adiabatically eliminate $\sigma^{x,y}_l$, which appear to be gapped throughout the whole parameter regime on the operator level. This results in an effective, coarse grained Heisenberg-Langevin equation for $n_l$. The final result is not different but the derivation more transparent than previous approaches.

The Heisenberg-Langevin equations for the spin operators $o_\alpha=\sigma^x_l, \sigma^y_l, n_l$ are operator valued stochastic differential equations 
\eq{eq12}{
\partial_t{o}_{\alpha}={\mathcal{D}}_{\alpha}+{\xi}_{\alpha}.
}
They contain a drift term ${\mathcal{D}}_{\alpha}$ and a quantum noise $\xi_{\alpha}$ \cite{Scully}. The drift ${\mathcal{D}}_\alpha$ describes the action of the adjoint superoperator on ${o}_\alpha$,
\eq{eq13}{
\mathcal{D}_{\alpha}=\mathcal{L}^* o_\alpha=i[H,o_\alpha]+\gamma\sum_l\left(\sigma^+_lo_\alpha \sigma^-_l-\frac{1}{2}\{n_l,o_\alpha\}
\right),
}

The quantum noise $\xi_{\alpha}$ describes the evolution of fluctuations of $o_\alpha$. It has zero mean $\langle \xi_{\alpha}\rangle_{\text{noise}}=0$, and an operator valued noise kernel $\chi_{\alpha\beta}=\langle \xi_{\alpha}\xi_{\beta}\rangle_{\text{noise}}$, where the noise average has to be understood as the average over all bath degrees of freedom (cf. \cite{Scully}).  It is determined via the Einstein relation \footnote{One should note, that $\chi_{\alpha\beta}$ can be equally well derived from a microscopic Hamiltonian, in which the system degrees of freedom are coupled to an appropriate bath that is responsible for the driven open nature of the system. For the present setup, this has been outlined in Ref.~\cite{NoiseFRGI}. Here, we consider the same setup but focus less on the microscopic implementation and thus refer to the Einstein relation \eqref{eq16} instead.}
\eq{eq16}{
\chi_{\alpha\beta}&=\langle\partial_t\left(o_\alpha o_\beta\right)-\left(\left(\partial_to_\alpha\right)o_\beta+o_\alpha\partial_to_\beta\right)\rangle_{\text{noise}}\nonumber\\
&=\langle\mathcal{L}^*\left(o_\alpha o_\beta\right)-\left(\mathcal{D}_\alpha o_\beta+o_\alpha\mathcal{D}_\beta\right)\rangle_{\text{noise}}.
}
The operator expectation value $\langle o_\alpha\rangle$ is then defined as the combination of noise average and system average, i.e. 
\eq{17}{
\langle o_\alpha\rangle=\text{Tr}\left(\rho\langle o_\alpha\rangle_{\text{noise}}\right).
}

The Heisenberg-Langevin equations for the local spin operators read as
\eq{eq18}{
\partial_tn_l&=-\gamma n_l +\Omega\sigma^y_l\Pi_l+\xi^n_l,\\
\partial_t\sigma^x_l&=-\Omega(\sigma^y_l s_l+s_l\sigma^y_l)-\frac{\gamma}{2}\sigma^x_l+\xi^x_l,\label{eq19}\\
\partial_t\sigma^y_l&=\Omega(\sigma^x_ls_l+s_l\sigma^x_l)-\frac{\gamma}{2}\sigma^y_l+\xi^y_l-2\Omega\left(2n_l-1\right)\Pi_l.\label{eq20}
}
We introduced the operator $s_l=\sum_{m \text{ nn } l}\left(\Pi_m-1\right)\sigma^x_m$. The hermitian noise operators $\xi^{n,x,y}_l$ have a flat power spectrum, i.e. they are $\delta$-correlated in time, due to the Markovian nature of the master equation. 

Defining the noise vector $\xif_l=(\xi^n_l, \xi^x_l, \xi^y_l)$ on each lattice site and the noise kernel $\chi_{l,m}=\langle \xif^\dagger_l \xif^{\phantom{\dagger}}_m \rangle_{\text{noise}}$ one obtains from the Einstein relation \eqref{eq16} that
\eq{eq21}{
\chi_{l,m}=\delta_{l,m}\left(\begin{array}{ccc}
\gamma n_l& \gamma\sigma^+_l& 
-i\gamma\sigma^+_l
\\ \gamma\sigma^-_l& \gamma & -i\gamma\\ i\gamma\sigma^-_l& i\gamma&\gamma
\end{array}\right).
}

The presence of a dark state is reflected in the Heiseberg-Langevin equations. For average values $\langle n_l\rangle=0$ for all $l$, the projectors $\langle \Pi_l\rangle=0$ vanish equally well and thus the lattice sites decouple. This implies $\langle \sigma^{\pm,x,y}_l\rangle=0$ for all $l$, such that the deterministic part of  Eqs.~\eqref{eq18}-\eqref{eq21} vanishes and the noise kernel $\chi_{l,m}$ is only non-zero in the $x-y$ sector. The latter ensures conservation of the spin algebra $\langle\sigma^x_l\sigma^y_l-\sigma^y_l\sigma^x_l\rangle=2i\langle\sigma^z_l\rangle=-2i$, $\langle\sigma^x_l\sigma^x_l\rangle=\langle\sigma^y_l\sigma^y_l\rangle=1$. This demonstrates that the property $\langle n_l\rangle=0, \forall l$,
\begin{itemize}
\item is always a possible solution for the Heisenberg-Langevin equations \eqref{eq18}-\eqref{eq21},
\item is the necessary and sufficient condition for ending up in the dark state $\rho_{DS}=|\psi_{\downarrow}\rangle\langle\psi_\downarrow |$,
\item leads to the absence of any density fluctuations in Heisenberg-Langevin equations. 
\end{itemize}

The densities $n_l$ thus represent the order parameters for the transition from a fluctuating, mixed steady state  ($n_l\neq0$) towards a fluctuation-less dark state ($n_l=0$). The fact that such a transition cannot be described in terms of an effective equilibrium statistical mechanics approach is reflected in the Heisenberg-Langevin framework by the multiplicative density noise $\xi^n_l\sim \sqrt{n_l}$. In the vicinity of $n_l=0$ this cannot be mapped to an effective, temperature-like noise kernel.

In order to derive an effective Heisenberg-Langevin equation for the density alone, we follow the common procedure and adiabatically eliminate the fast variables $\sigma^{x,y}_l$ from the set of equations \eqref{eq18}-\eqref{eq20}. 
After some algebra and the restriction to nearest neighbor couplings, one finds the operator equation for the spin-up density
\eq{eq29}{
\partial_tn_l=-\gamma n_l-2\Omega^2\Pi_l(\mathcal{K}^{-1})_{l,j}(2n_j-1)\Pi_j+\tilde{\xi}^n_l,
}
with the (regular) retarded operator 
\eq{eq28}{
\mathcal{K}_{l,j}=\delta_{l,j}\tfrac{\gamma}{2}+\delta_{j,\text{nn }l}\frac{32\Omega^2}{\gamma^2}(1-\Pi_j)+i0^+
}
and an effective noise $\tilde{\xi}^n_l$ with kernel
\eq{eq30}{
\langle \xi^n_l\xi^n_m\rangle_{\text{noise}}=\delta_{l,m}\gamma n_l+\gamma\Omega^2\Pi_l(\mathcal{K}^{-2})_{l,m}\Pi_m.
}
These equations are valid on time scales $t>\gamma^{-1}$, which is the decay scale of $\sigma^{x,y}_l$ such that these operators could be treated effectively static (see ~\cite{SM} for details).

In order to analyze the dynamics described by the Heisenberg-Langevin equation for the operators $n_l$  \eqref{eq29} with the noise kernel \eqref{eq30}, we perform a long-wavelength analysis of the corresponding Langevin equation for the coarse grained density expectation value $\nu(x,t)$, where $x$ is a continuous variable in $d$-dimensional space and $t$ is time. We define $\nu(x,t)$ as the excitation density in a $d$-dimensional, small but macroscopic volume, which contains $N_x\gg 1$ lattice sites centered around the coordinate $x$, i.e. 
\eq{eq31}{
\nu(x,t)=\frac{1}{N_x a^d}\sum_{j=1}^{N_x}\langle n_{j}\rangle(t).
}

The Langevin equation for $\nu_X=\nu(x,t)$ is obtained by evaluating the operator expectation on the R.H.S. of Eq.~\eqref{eq29} within three crucial approximations:
\begin{itemize}
\item A site decoupling mean-field approach, i.e. $\langle n_j n_l\rangle(t)=\langle n_j\rangle(t)\langle n_l\rangle(t)$ for different sites $l\neq j$. 
\item The temporal decoupling on identical lattice sites, i.e. $\langle n_ln_l\rangle=\langle n_l\rangle(t)\langle n_l\rangle(t')$ if $t$ and $t'$ are not identical. This is important in order to evaluate the product $\Pi\tilde{\mathcal{K}}^{-1}\Pi$, for which $\tilde{\mathcal{K}}^{-1}$ has to be understood as an infinitesimal (retarded) time evolution operator, which evolves $\Pi$ from $t$ to $t+\epsilon$. 
\item The derivative expansion of $\Pi_l$, incorporating at most second order derivatives, i.e. $\langle\Pi_l\rangle\rightarrow 2d\nu_X+a^2 \nabla^2 \nu_X+d(2d-1)\nu^2_X+o(\nabla^4\nu, (\nabla^2\nu)^2)$.
\end{itemize}
This yields the fundamental Langevin equation of our approach (prime denotes derivative w.r.t. the argument)
\eq{eq32}{
\partial_t\nu_X=D\nabla^2\nu_X-V'(\nu_X)+\xi_X.
}
It describes the diffusive propagation ($D=\Omega^2a^2/\gamma$) of the density in the effective, local potential
\eq{eq35}{
V(\nu_X)=\frac{\Delta}{2}\nu_X^2+\frac{\mu}{3}\nu^3_X+\frac{\lambda}{4}\nu^4_X
}
subject to a non-thermal, multiplicative noise with kernel
\eq{eq33}{
\langle \xi_X\xi_Y\rangle=\delta(X-Y)\kappa \nu_X,
}
whose dynamics will be investigated in detail in the following sections.
Neglecting the renormalization under coarse graining, the effective parameters in the Langevin equation are related to the master equation via
\eq{eq34}{
\frac{\Omega^2}{2\gamma}+\Delta=\kappa=\gamma , \ \lambda=\frac{4 \Omega ^2 d(2d+1)}{\gamma},\ \mu=\frac{-2\lambda d}{2d+1}.
}

\section{Scope \& Methods}
\label{sec:droplets}
The Langevin equation~\eqref{eq32} encodes the complete physical content of the dark state transition. Its (numerical) solution is, however, only rarely a viable approach to resolving the full dynamics of the model. Aside from computational cost in higher dimensions, the numerical realization of a multiplicative noise itself is a challenging and so far not unambiguously solved problem \cite{Dickman1994,Munoz2005,Yuhai1997,SIAM2009,wiktorsson2001}. Therefore, we rely on a more diverse assortment of methods to develop an understanding of the phenomenology. 

In order to familiarize the reader with the model, we start with a brief discussion of the corresponding nonequilibrium path integral in the saddle-point approximation. This repeats basic elements from the analysis in Refs.~\cite{Marcuzzi2016,NoiseFRGI} and serves as an introduction to the phase diagram and to the notion of a first order dark state transition. It also gives rise to a number of questions to be addressed in this work.
 
We then develop a functional renormalization group (fRG) framework that can be applied in arbitrary dimensions and includes spatial fluctuations and noise beyond the mean field picture. The approximate nature of this approach calls for cross-validation, which we perform with the help of Langevin simulations in one dimension.

\subsection{Path integral formulation and mean-field phase diagram}
Any dynamics described in terms of a stochastic Langevin equation can be mapped onto a corresponding nonequilibrium path integral via the Martin-Siggia-Rose-Janssen-de Dominicis (MSRJD) construction \cite{MSR,MSR+J,MSR+D}. This allows us to approach the theory via suitable functional methods such as saddle point equations and the renormalization group. The MSRJD construction formally computes the partition function by summing over all stochastic trajectories (see, e.g.,~\cite{Kamenev} for details). Introducing the purely imaginary, so-called response field $\tilde{\nu}_X$ it reads 
\begin{equation}
\label{MSRInt}
\mathcal{Z} = \int\mathcal{D}[\nu,\tilde{\nu}] \exp\left(-S[\nu,\tilde{\nu}] \right).
\end{equation}
With the corresponding action
\eq{MSRS}{
S[\nu,\tilde{\nu}] = \int_X\tilde{\nu}_X\left[\partial_t\nu_X - D\nabla^2\nu_X + V'(\nu_X) - \kappa\tilde{\nu}_X\right].
}
It is at most quadratic in the response fields $\tilde{\nu}_X$, which is a consequence of the ferromagnetic dark state. This state displays no quantum fluctuations in the order parameter $n_l$ such that higher order noise terms, familiar from the \emph{Keldysh} quantum path integral \cite{SiebererReview,Kamenev}, become subleading in the limit $\nu_X\rightarrow0$. Within our analysis, we validated that even at non-zero $\nu_X$ these terms do not qualitatively alter the picture obtained with the renormalization group and thus can be neglected right from the start.

\begin{figure}[t]
\centering
\includegraphics[width = 8cm]{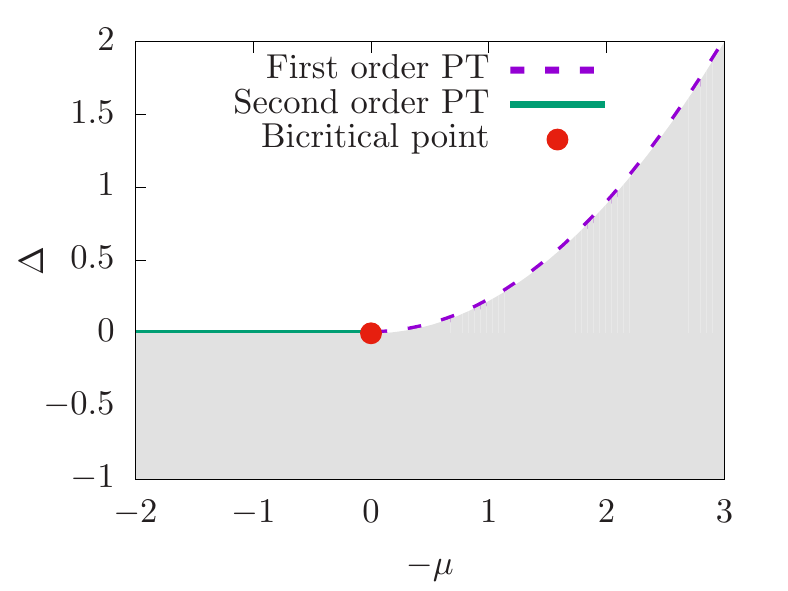}
\caption{Mean-field phase diagram for the action~\eqref{MSRS} at $\lambda =1$. The finite-density phase is indicated by shading. A bicritical point at $\mu=0$ separates regimes of first (dashed line) and second order (solid line) phase transitions.}
\label{MFPD}
\end{figure}

We will now briefly review the saddle-point analysis of~\cite{NoiseFRGI}. In the absence of spatial and temporal fluctuations ($\nu_X=\nu$) one finds the saddle-point equations
\eq{MFEqs}{
\begin{aligned}
\fdif{S}{\nu_X} \overset{!}=0\quad \Leftrightarrow \quad & \tilde{\nu} V''(\nu) -\kappa \tilde{\nu}^2= 0,\\
\fdif{S}{\tilde{\nu}_X}  \overset{!}=0\quad \Leftrightarrow \quad & V'(\nu) -2 \kappa\tilde{\nu}\nu = 0.\\
\end{aligned}
}
The stationary field configurations are considered to be insensitive to the noise level ($\tilde{\nu} = 0$) and the phase diagram is derived from the deterministic part of the action only:
\eq{DetMF}{
0 \overset{!}{=} V'(\nu) = \Delta\nu + \mu\nu^2 + \lambda\nu^3.
}
Keeping $\lambda>0$ fixed, the phase diagram in Fig.~\ref{MFPD} ensues.
In general, the deterministic potential $V$ admits two qualitatively different states: a finite-density (active) state at $\nu_f = \frac{\sqrt{\mu^2 - 4\Delta\lambda} - \mu^2}{2\lambda}$ and the (inactive) dark state at $\nu_d=0$. For $\Delta < 0$, $\nu_f$ is the only minimum of $V$, whereas for $\Delta > 0$, $\mu < 0$ both are present. Consequently, the transition between active and inactive phases as a function $\Delta$ is of second order for $\mu\geq 0$, whereas a first order transition occurs for $\mu < 0$.

The regime displaying a second order phase transition, including nonequilibrium universality and scaling, was investigated in detail in Ref.~ \cite{NoiseFRGI}. Here we focus on the domain of first-order phase transitions ($\Delta > 0$, ${\mu > 0}$). As opposed to second order transitions, there is no universality associated with this type of phase boundary. Consequently, a different set of questions poses itself and shall be answered during the course of this work:
\begin{itemize}
\item Can the notion of a first order dark state transition persist beyond the saddle-point approximation and if so, how sensitive is it with respect to spatial dimensionality?
\item What distinguishes a first order dark state transition from its thermal counterpart?
\item Is there a similar notion of phase coexistence for dark states?
\item Can we shed light on the dynamics at multiple scales by adapting (functional) renormalization group techniques?
\end{itemize}

\subsection{Renormalization Group I: Full Potential Approximation}
\label{RGTech}
At this point, it is not clear whether the mean-field phase diagram in Fig.~\ref{MFPD} is at all a faithful representation of the system's physical behavior. The multiplicative noise, being the signature feature of the model, has been omitted completely so far. In previous work, this has partially been remedied by employing an optimal path approximation~\cite{NoiseFRGI}. However, this approach does not account for corrections to the noise itself due to the interplay of deterministic dynamics and fluctuations. As we will point out later, such corrections can in fact be very sizable. Here, we therefore develop an fRG scheme capable of implementing precisely this.

In previous work~\cite{Canet2004,Canet2006,Canet2005,Buchhold2016,NoiseFRGI}, (functional) RG techniques have successfully been applied to investigate critical phenomena in the presence of a dark state. Their unique advantage is a systematic inclusion of fluctuation effects beyond the noiseless mean-field or optimal path~\cite{Kamenev} approximations while being rather inexpensive compared to full-scale numerical simulations. In the following, we will therefore extend the fRG scheme from~\cite{NoiseFRGI} in order to accommodate an investigation of discontinuous phase transitions.

The main objective of a functional renormalization group analysis is the computation of the effective action $\Gamma$, i.e. the generating functional of one particle irreducible correlation and response functions. This is achieved by solving the Wetterich equation~\cite{Wetterich1993}
\begin{equation}
\label{eq:Wetterich}
\partial_k \Gamma_k = \frac{1}{2}\mathrm{Tr}\left[(\partial_k R_k)\left(\Gamma_k^{(2)} + R_k\right)^{-1}\right].
\end{equation}
Here, $\Gamma_k$ is the so-called effective average action, interpolating between the microscopic $S = \Gamma_{k\rightarrow\Lambda}$ and the full effective $\Gamma = \Gamma_{k=0}$ actions. This is achieved by means of an (additive) regulator function $R_k$ that introduces a dependence on a scale parameter $k \in [0,\Lambda]$, with $\Lambda$ the overall energy/momentum cutoff.

While Eq.~\eqref{eq:Wetterich} by itself is an exact representation of the underlying field theory, its solution for nontrivial systems generally requires approximations. As $\Gamma_k$ does in principle facilitate all terms compatible with the symmetries of the respective system, approximations are invoked by reducing the number of terms kept during the calculation down to a manageable number. Thus, a truncated ansatz for $\Gamma_k$ is constructed, whose second functional derivative $\Gamma_k^{(2)}$ may then be used to compute the right hand side of Eq.~\eqref{eq:Wetterich} and determine the evolution of the action.

The crucial point in obtaining reliable and accurate results is of course the very choice of terms to be kept in $\Gamma_k$. While canonically relevant or marginal operators are an obvious choice for weakly coupled systems (i.e. in the vicinity of a gaussian fixed point), the situation is less clear if one is interested in the physics of interacting fixed points or even non-universal regimes. The latter case is of particular interest for this work, as it encompasses discontinuous phase transitions. 

Unfortunately, it has so far not been possible to devise a construction scheme for truncated $\Gamma_k$ that guarantees a steady improvement of results upon inclusion of further contributions. We therefore resort to the well-established derivative expansion~\cite{Berges2002} and benchmark our results against other methods whenever possible.

The generic ansatz for $\Gamma_k$ based on the action~\eqref{MSRS} at lowest non-trivial order in the derivative expansion is
\begin{equation}
\label{eqGkBloat}
\Gamma_k = \int_X \tilde{\nu}_X\left[Z_k \partial_t + D_k \nabla^2\right]\nu_X + \int_X W(\tilde{\nu}_X,\nu_X).
\end{equation}
Here, $W_k(\tilde{\nu}_x,\nu_X)$ is the \emph{local potential} that contains arbitrary (symmetry compatible) powers of $\tilde{\nu}_X$, $\nu_X$ and products thereof. Its initial value according to Eq.~\eqref{MSRS} is given by
\begin{equation}
\label{eq:InitialW}
W_\Lambda(\tilde{\nu}_X,\nu_X) = \tilde{\nu}_X\left[\Delta\nu_X + \mu\nu_X^2 + \lambda\nu_X^3 - \kappa\tilde{\nu}_X\right].
\end{equation}

While it is straightforward algebra to determine the flow equation for $W(\tilde{\nu}_X,\nu_X)$, its actual evaluation cannot be done analytically anymore. Even numerically, it is a rather costly task, as it requires the discretization of $\nu_X$ and $\tilde{\nu}_X$ on a sufficiently fine-grained two-dimensional grid. 

Considering Simplifications, the different canonical scaling and thus relevance of field monomials $\tilde{\nu}_X^{m}\nu_X^n$ seems to justify a representation of $W_\Lambda(\tilde{\nu}_X,\nu_X)$ as a polynomial. However, this is not feasible in our situation for two reasons.

First, the deterministic potential in the vicinity of a discontinuous phase transition exhibits two competing minima even at mean-field level. Due to their finite separation in field space, a polynomial expansion of $V_k(\nu_X)$ about one minimum therefore becomes a rather bad approximation in the vicinity of the other one. since the difference of the minimas values is crucial for determining the phase boundary, a low-order polynomial expansion will not yield accurate results~\cite{Boettcher2015,Roscher2015}.

Second, it can be shown~\cite{Berges2002} the $V_k(\nu_X)$ is rendered \emph{convex} for $k\rightarrow 0$. For the effective potential at a first-order phase transition, this amounts to a Maxwell construction. The ensuing straight line is represented at best asymptotically by a polynomial expansion even at intermediate scales. In addition, we demonstrate that the initial linear multiplicative noise is modified drastically and cannot be represented by a simple expansion. 

We therefore refrain from a polynomial representation in both field variables and keep the full $\nu_X$ dependence on the deterministic potential as well as the noise kernel:
\begin{equation}
\label{eq:WAcTrunc}
W_k(\tilde{\nu}_X,\nu_X) = \tilde{\nu}_X\left[u_k(\nu_X) + \chi_k(\nu_X)\tilde{\nu}_X\right],
\end{equation}
with $u_k(\nu_X) \equiv V_k'(\nu_X)$ and the initial conditions indicated in Eq.~\eqref{eq:InitialW} above. We checked numerically that higher order terms $\sim\tilde{\nu}_X^3$ as naturally present in a Keldysh framework do not contribute qualitatively and give only small corrections to quantitative results. On similar grounds, we set the wavefunction renormalization parameters $Z_k = D_k = 1$ from now on.

The flow equations for the (derivative of the) deterministic potential and the noise kernel can now be computed:
\begin{subequations}
\label{DSFlowEqs}
\begin{align}
\partial_k u_k =& -\frac{k^{d+1}v_d\chi_k u''_k}{d\left(k^2+{u'_k}^2\right)^2}, \\
\partial_k \chi_k =& -\frac{k^{d+1}v_d\chi_k}{2d\left(k^2+{u'_k}^2\right)^4}\cdot\left[3\chi_k{u''_k}^2 \right.\\
&-\left. 8\left(k^2+{u'_k}^2\right)\chi'_ku''_k + 2\left(k^2+{u'_k}^2\right)^2\chi''_k \right]\nonumber.
\end{align}
\end{subequations}
Here, $v_d = \left[2^{d-1}\pi^{\frac{d}{2}}\Gamma\left(\frac{d}{2}\right)\right]^{-1}$ is the (normalized) volume of the $(d-1)$-dimensional unit sphere.

\subsection{Renormalization Group II: Benchmarking}
For (nonequilibrium) systems with mean-field first order phase transitions, there is neither literature (cf. \cite{Berges1997I,Jakub2009,Delamotte2004} for equilibrium approaches) nor a stringent analytical argument for the sufficiency of the truncation scheme presented above. We therefore need to benchmark the results of our fRG computations in order to build confidence in the reliability of its predictions. In order to achieve this, we conducted numerical simulations of the Langevin equation~\eqref{eq32} in one spatial dimension. In the following, we will present two different physical setups and compare the results of these simulations to the respective outcomes of our fRG approach.

\subsubsection{Thermal equilibrium}
A simple test case is provided by the thermal equilibrium limit. Since we consider the one-dimensional case with short-range interactions, no phase transitions can be expected due to the exponential growth of entropy with number of domain walls \cite{LanLif}. Even if mean-field predicts the existence of first-order transitions, we expect the occurrence of a smooth crossover upon inclusion of fluctuations. 

\begin{figure}[t]
\centering
\includegraphics[width = 0.5\textwidth]{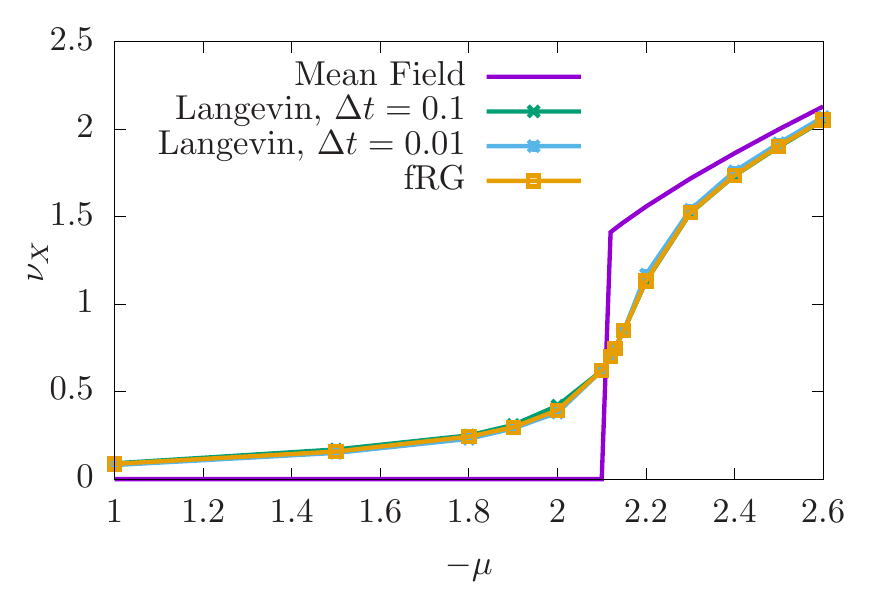}
\caption{Density $\nu_X$ as a function of $\mu$ for $\Delta = 1$, $\lambda = 1$ at $T=0.25$. Langevin and fRG results agree quantitatively, whereas sizable corrections to the mean-field prediction are made. Statistical error bars on the Langevin results are not visible on the scale of the plot.}
\label{ThermLang}
\end{figure}

For an appropriate implementation in the Langevin equation~\eqref{eq32}, the noise correlation~\eqref{eq33} has to be replaced by
\begin{equation}
\label{ThermNoise}
\langle\xi_X\xi_Y\rangle = 2T\delta(X-Y),
\end{equation}
with $T$ being the temperature. Our simulations were performed on a 1000 site lattice with periodic boundary conditions and for time steps of $\Delta t = 10^{-1}...10^{-2}$ in units of the lattice spacing. We picked a regime that encloses a first order phase transition at mean-field, given by $\Delta = 1$, $\lambda = 1$ at $T=0.25$. The outcomes for the average density $\nu_X$ as a function of $\mu$ are shown in Fig.~\ref{ThermLang} alongside the corresponding mean-field and fRG results. For the functional RG calculation, we set
\begin{equation}
\chi_\Lambda(\nu_X) = \chi_k(\nu_X) = T,
\end{equation}
in order to implement detailed balance.
Thus, our truncation is simplified further by not allowing for any feedback of the deterministically induced fluctuations into the flow of the noise kernel.
This approximation can only be expected to be appropriate at sufficiently high temperatures. Since $\partial_k u_k \sim T$, it is obvious that the flow would be switched off for $T\rightarrow 0$. This situation therefore corresponds to a purely classical setup, where only thermal fluctuations are taken into account, whereas quantum fluctuations are neglected. Judging by the results presented in Fig.~\ref{ThermLang}, functional RG is very well able to reproduce this limit, as the curves for the Langevin simulation lie practically on top of the fRG result.

\subsubsection{Non-equilibrium: dark state}
The thermal limit mainly constitutes a test for the deterministic sector of the model. Naturally, we need to benchmark our fRG also in a nonequilibrium situation, i.e. in the presence of non-trivial noise. We therefore implemented the Langevin equation~\eqref{eq32} with the original multiplicative noise~\eqref{eq33} as well. Unfortunately, this is not as straightforward as in the thermal case. In particular, the concrete realization of the multiplicative noise in a time-discrete setting is still an open problem that has not been solved unambiguously \cite{SIAM2009,wiktorsson2001}. One basic issue is that upon discretization, a multiplicative noise is able to induce negative values for the density where it is itself not well defined anymore \cite{Dickman1994,Munoz2005,Yuhai1997}. While no negative density can be observed since the respective fluctuations disappear on average, the definition of the noise kernel for negative densities does have an impact on the actual value of observables.

\begin{figure}[t]
\centering
\includegraphics[width = 0.5\textwidth]{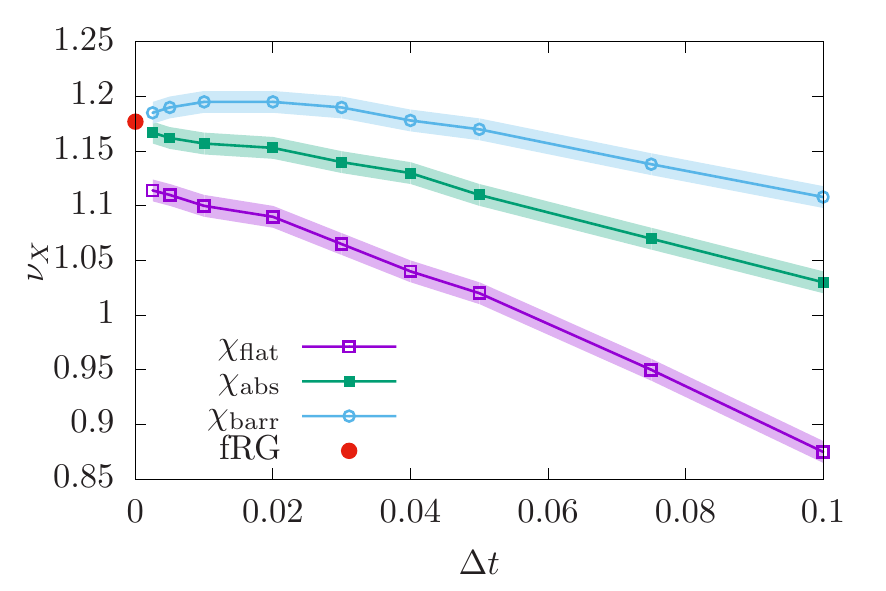}
\caption{Density $\nu_X$ at different temporal discretizations for constant $N_t\cdot \Delta t = 100$ at $\Delta = 1$, $\mu = -2.42$, $\lambda = 1$ and $\kappa = 0.5$. Statistical errors on the Langevin results are depicted by the shading around the data points. Approximate convergence towards the fRG results can be observed for dark state realizations $\chi_{\rm{abs}}$ and $\chi_{\rm{barr}}$, cf. Eqns.~\eqref{DarkNum}.}
\label{DSLangSC}
\end{figure}

We employ three different definitions of the noise kernel for negative $\nu_X$:
\begin{subequations}
\label{DarkNum}
\begin{align}
\chi_{\rm{flat}} &= \left\{ \begin{matrix} \kappa \nu_X, & \nu_X \geq 0 \\ 0 , & \nu_X < 0 \end{matrix}\,\,, \right.\\
\chi_{\rm{abs}} &= \kappa|\nu_X|, \\
\chi_{\rm{barr}} &= \left\{ \begin{matrix} \kappa \nu_X, & \nu_X \geq 0 \\ \infty , & \nu_X < 0 \end{matrix}\,\,. \right.
\end{align}
\end{subequations}
While the resolution of the temporal lattice did not play a big role in thermal equilibrium (cf. Fig.~\ref{ThermLang}), this is not true for the present setup anymore. An extrapolation to $\Delta t = 0$ would therefore be necessary for a quantitative comparison with fRG results. Due to the increased numerical effort necessary to achieve reasonable accuracy, and since our main focus is not on Langevin simulations, we refrain from doing so. In Fig.~\ref{DSLangSC}, results for different $\Delta t$ and all three realizations of the noise kernel are presented in comparison with the fRG result.
The configuration belonging to these data points is given by the (initial) values $\Delta = 1$, $\mu = -2.42$, $\lambda = 1$ and $\kappa = 0.5$ on a grid of 100 sites. This setup lies well inside of a finite-density phase at mean-field and also according to fRG results (see Fig.~\ref{1DPD} below). For comparability, the number of steps times the temporal spacing $N_t\cdot \Delta t = 100$ have been kept constant. Each single data point results from an average over 1200 sample densities. Statistical uncertainties are indicated by shading.
While the estimate obtained with $\chi_{\rm{flat}}$ is somewhat low, $\chi_{\rm{abs}}$  as well as $\chi_{\rm{barr}}$ are well compatible with the fRG result. 

For completeness, it should be stated that convergence towards the continuous time and infinite volume limit is not achieved as easily in low-density regimes. While $N = 100$ sites appeared to be sufficient for the setup presented in Fig.~\ref{DSLangSC}, this is not the case for smaller $|\mu|$ anymore. The reason is once again the difficulty of resolving the actual dark state in the Langevin simulations. Again, we refrained from a more sophisticated numerical analysis of the Langevin equation in this regime due to numerical cost. However, it can be stated that the position of tentative phase transitions appears to be compatible with the fRG result. The agreement is even better, if the flow equations~\eqref{DSFlowEqs} are supplemented by a term $w_k(\nu_X)\tilde{\nu}_X^3$. No qualitatively notable differences were found, though, and the quantitative corrections are on the percent level. We therefore do not include $w_k(\nu_X)$ into our further analyses.

In conclusion, we found good agreement between numerical Langevin simulations and our functional RG results in and out of equilibrium. This gives us confidence that we can trust our fRG approach for an extended analysis of the one-dimensional as well as the three-dimensional cases.

\begin{figure*}
\centering
\includegraphics[width=0.8\textwidth]{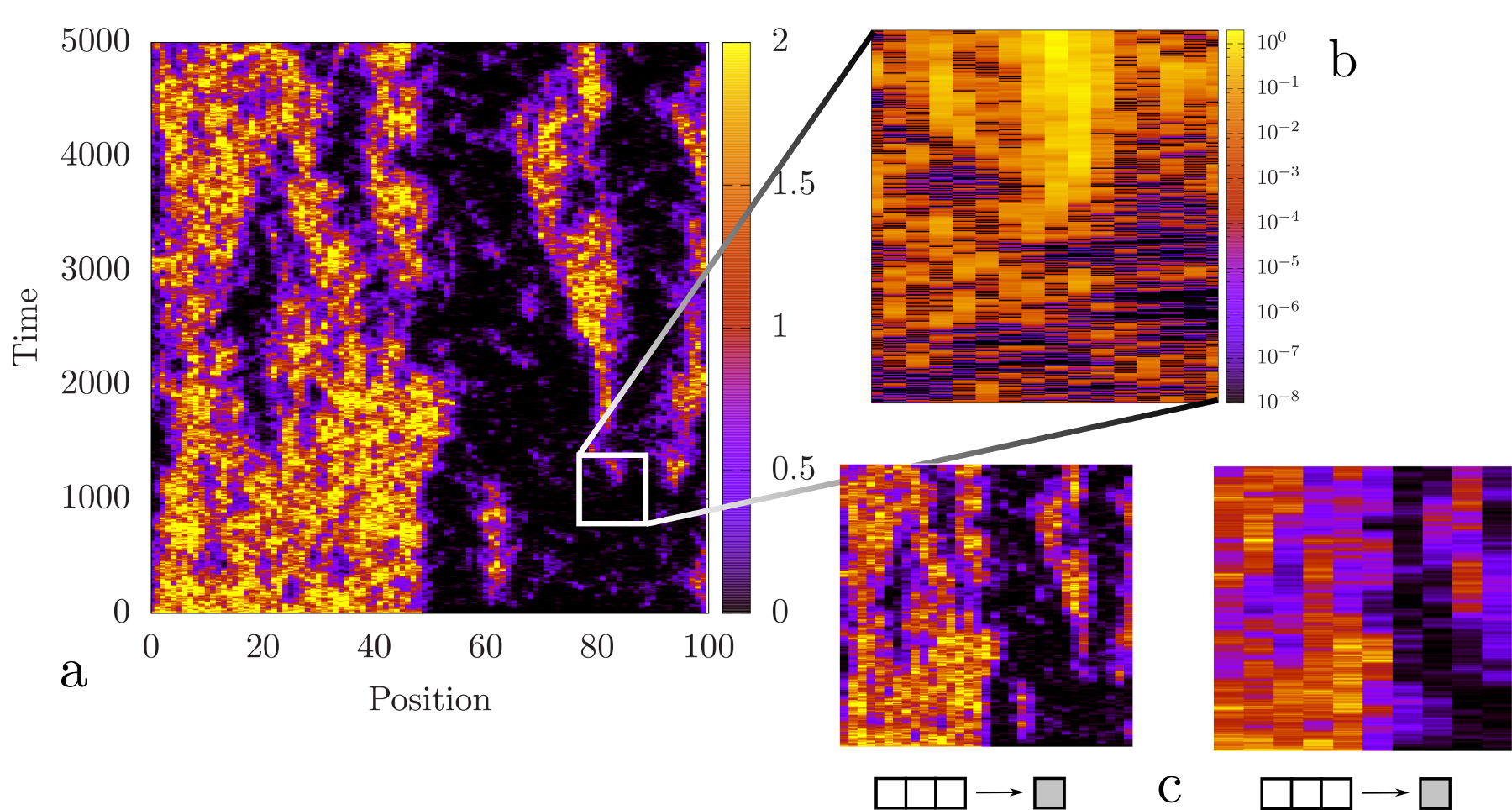}
\caption{Evolution of the density field $\nu_X$ as a function of position and time for parameters very close to the dark state phase transition. The initial configuration is a step function $\nu_{x,t=0}=\nu_0\Theta(L/2-x)$, where $L=100$ is the system size and $\phi_0$ is the value of the finite density potential minimum. a) Local clusters of nearly vanishing density are spontaneously formed in the region with $x<L/2$ and establish extended, noisy interfaces with the finite density regions. a+b) Exponentially small density fractions diffuse into the initial dark state regime ($x>L/2$) and spontaneously form clusters of finite density, again with noisy and finite interface. c) The noisy character of the interfaces is not diminished upon coarse graining, hinting at the emerging scale invariance.}
\label{figLangevin}
\end{figure*}

\section{The fate of the dark state in one dimension}
\label{sec1d}
Including both spatial and temporal fluctuations beyond the mean-field approximation often has dramatic consequences on the stability of ordered phases and the nature of the corresponding phase transitions. This is most drastically expressed in one dimension, where in the presence of a finite noise level the entropy gain per excitation always wins against the energy cost \cite{LanLif} and no ordered phase can be stabilized (see Fig.~\ref{ThermLang}). Spatial correlations in one dimension are so strongly pronounced, that even infinitesimally small temporal fluctuations of the order parameter can grow rapidly in space and render the ordered state unstable. While the dark state itself does not experience any fluctuations and is thus immune to this destructive mechanism, any infinitesimal deviation from a pure dark state does experience strongly suppressed but finite fluctuations and it is thus a priori unclear whether a dark state phase can persist against the generically strong fluctuation dynamics in one dimension.

Here, we analyze the fate of the dark state in the presence of fluctuations. We apply the functional renormalization group approach introduced above, which includes temporal and spatial fluctuations on all energy and momentum scales, and numerical simulations of the corresponding Langevin equation. From the combination of both approaches, we obtain a comprehensive picture of the asymptotic dynamics in the regime where both the dark state and the fluctuating state are deterministically stable on the mean-field level. 

From our analysis, we conclude that
\begin{itemize}
\item Fluctuations render one of the two mean-field steady states unstable and only one, unique steady state remains in the asymptotic time limit. 
\item The dark state phase persists in the presence of spatial and temporal fluctuations and remains stable for intermediate and low potential barriers.
\item For large potential barriers, the dark state becomes unstable and decays into a mixed state with non-vanishing field expectation value. 
\item Fluctuations {\bf enlarge} the regime of the dark state phase compared to the mean-field prediction but soften the predicted first order phase transition towards a continuous, {\bf second order transition} throughout the whole parameter regime.
\end{itemize}

The most striking result of these conclusions is that fluctuations are sufficiently strong to remove the mean-field predicted first order phase transition completely and replace it by a continuous second order phase transition throughout the entire one-dimensional parameter regime \cite{Jakub2009}. On the other hand, they are sufficiently weak that a dark state phase, which is represented by only a single fluctuationless point in phase space, can be stabilized for an extended parameter regime. Similar behavior has been observed for classical, absorbing state phase transitions in one spatial dimension, for which first order phase transitions induced by short ranged processes are conjectured not to exist \cite{Hinrichsen1or}, but second order absorbing state phase transitions have been established \cite{Janssen2005,DP_Hinrichsen}.

The observed softening of the transition, which is a major feature of the one-dimensional dynamics, can be understood in a two-staged coarse graining procedure. In order to do so, imagine one initializes the dynamics in the mixed phase but close to the phase border with some field configuration $\nu_X>0$. In addition to regular noise induced, small oscillations around the deterministic field expectation value (green configuration in Fig.~\ref{figPot}), some rare, strong noise kicks let the field climb up and eventually overcome the potential barrier between the mixed and the dark state. This leads to the formation of local clusters with exponentially small field configurations, see Fig.~\ref{figLangevin}. In one dimension, the interplay between noise and kinetic energy, however, leads to spatially extended, fluctuating interfaces between the dark state clusters and the finite-density phase, which spoil the picture of well-defined domain wall interfaces (red configuration in Fig.~\ref{figPot}). As a consequence, meta-stable saddle-point configurations, known as droplets or instantons, that interpolate smoothly between the two phases (blue configuration in Fig.~\ref{figPot}) are only observable on short times and become rather unstable at intermediate and larger times. This leads to a breakdown of the droplet picture in one dimension.

\begin{figure}
\centering
\includegraphics[width=0.5\textwidth]{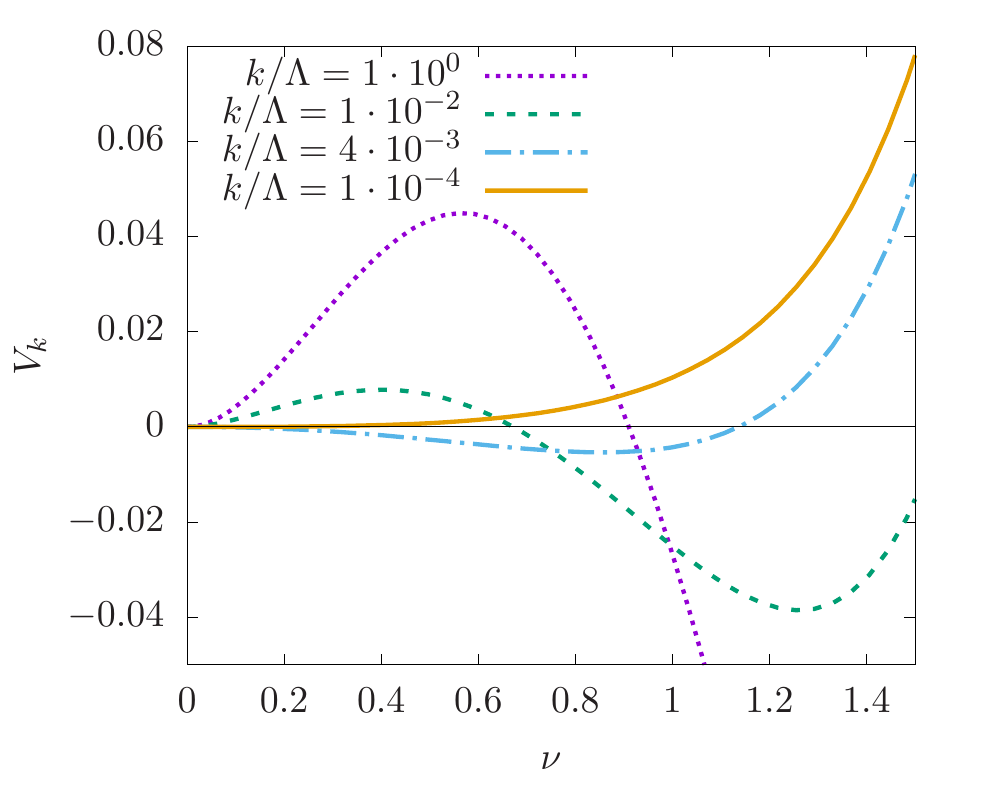}
\caption{Evolution of the deterministic potential $V_k$ for $\lambda = 1$, $\mu = -2.33$, $\Delta  =1$ and $\kappa = 0.5$. Before the inactive state becomes the only nontrivial minimum at $k\rightarrow0$ (solid line), the potential barrier is removed (dot-dashed line).}
\label{1DPotEvo}
\end{figure}

Upon coarse graining, the noisy interfaces between the two phases continuously smoothen the potential well separating the dark and the mixed state, which leads to a continuously varying density and, close to the phase transition, to the emergence of scale invariance on intermediate and large distances. This is observed in the numerical simulations of the Langevin equation, see Fig.~\ref{figLangevin}, as well as in the numerical solution of the fRG flow equation. Within the latter, the potential barrier separating the dark and the mixed state vanishes smoothly while short distance modes are integrated out, see Fig.~\ref{1DPotEvo}. Already on intermediate scales, the potential $V_k$ develops a single minimum and becomes convex, indicating the absence of sharp domain walls between dark and mixed states~\cite{Note1}.

This observation, quantitatively and qualitatively, resembles the dynamics close to a second order phase transition in the presence of a nonequilibrium noise. Indeed, increasing the noise strength, the Langevin simulations display a smoothly vanishing order parameter expectation value instead of the predicted, first order jump. Within the fRG framework, the corresponding observation is that on intermediate scales the potential barrier fades away, leading to a single deterministic minimum and the breakdown of bistable behavior, see Fig.~\ref{1DPotEvo}. Continuing coarse graining to the largest scales, integration over the long distance modes lets this minimum continuously move towards zero field expectation value, i.e. the dark state. At the transition, the minimum reaches $\nu = 0$ at asymptotically small scales. In the RG interpretation, this describes the continuous divergence of the correlation length, i.e. the emergence of scale invariance as the transition is approached. The removal of the barrier during the flow with major corrections to the position of the nontrivial minimum still ahead also provides an intuitive explanation for the failure of the droplet picture: in a convex potential with only one minimum, no metastable droplet solution can be defined anymore. Successively integrating out fluctuations of larger and larger length scales in the vicinity of the phase transition, we find that the barrier always vanishes before the non-trivial minimum itself inside the inactive phase~\cite{Note1}. While droplet-like fluctuations therefore do play a role at short length scales, the fate of the minimum is ultimately decided by complex field configurations such as indicated by the red line in Fig.~\ref{figPot}.

\begin{figure}
\centering
\includegraphics[width=0.5\textwidth]{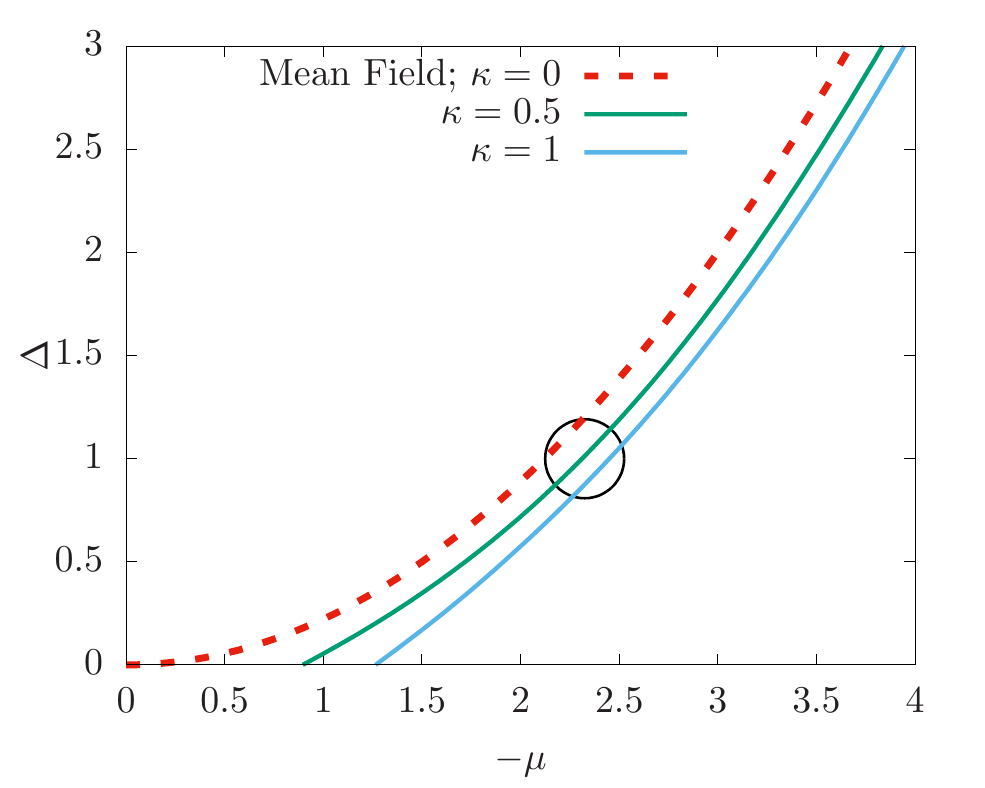}
\caption{Phase diagram of the one-dimensional system in the $\Delta-\mu$ plane for fixed $\lambda=1$. The mean-field first order transition (dashed line) is changed to second order (solid lines) upon inclusion of fluctuations. Increased initial noise levels lead to a larger inactive phase. The region where Figs.~\ref{1DPotEvo} and~\ref{1DNoiseEvo} are situated is indicated by a circle.}
\label{1DPD}
\end{figure}

Another general consequence of the presence of fluctuations is the increased probability of the system to end up in the inactive phase, despite being in an active configuration on deterministic grounds alone. Thus, an increased initial noise level leads to an enlarged dark state phase, see Fig.~\ref{1DPD}.

While we have discussed the reasons for the existence of an inactive phase beyond mean field, one could also wonder about the stability of the finite-density phase in the presence of a dark state. Intuitively, there is always a possibility that some rare fluctuation drives the whole system into the zero-density dark state from which it cannot escape anymore. In fact, such events do plague Langevin simulations which are by construction bound to finite systems \cite{SIAM2009, wiktorsson2001}. The infrared endpoint of the RG evolution, on the other hand, provides access to the steady state itself. While \emph{fluctuation} effects are included which have to be accounted for by averaging over samples in a Langevin treatment, this state is not necessarily \emph{noiseless}: the noise kernel $\chi_k$ (and potentially higher order operators as well) do evolve alongside the deterministic potential $V_k$. They generally remain finite for $k\rightarrow 0$, i.e. even after all fluctuations are integrated out. The existence of a nontrivial minimum in $V_{k\rightarrow 0}$ in itself is therefore not yet sufficient to argue for the stability of the finite-density phase, as the noise $\chi_k$ has to be taken into account also in the deep infrared. 

\begin{figure}
\centering
\includegraphics[width=0.5\textwidth]{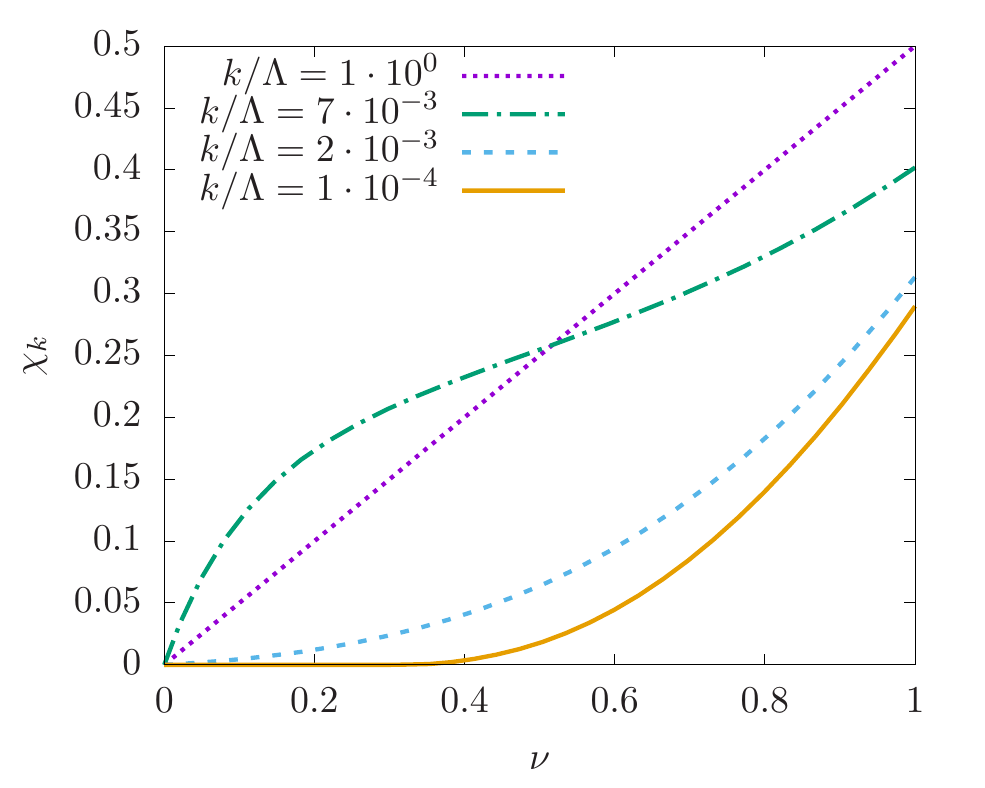}
\caption{Evolution of the noise kernel $\chi_k$ for $\lambda = 1$, $\mu = -2.35$, $\Delta  =1$ and $\kappa = 0.5$. At low $k$, the dark state spreads (solid line) and thus stabilizes the finite-density state.}
\label{1DNoiseEvo}
\end{figure}

Fig.~\ref{1DNoiseEvo} shows the evolution of the noise kernel for a configuration inside the finite-density phase. The crucial finding is that for $k\rightarrow 0$, the initially single (point-like) dark state at $\nu_X=0$ spreads into a continuum, i.e. to finite values of $\nu$. While it does not necessarily reach the value of the steady state $\nu_f$, it suppresses any noise between $\nu_f$ and the initial dark state at $\nu_X=0$. Since at the same time $u_{k\rightarrow 0}$ becomes convex and establishes a single minimum at $\nu_f$, the dark state becomes unstable and the coarse grained Langevin equation describes deterministic, noiseless motion towards a mixed steady state.

The universality class at this second order phase transition could not yet be determined. The computation times for a sufficiently clear resolution of the critical exponents in the limit $k\rightarrow0$ are currently requiring too many resources and one possibly has to think about another approach to detect the critical exponents. We want to stress here, that both the noise kernel $\chi_k$ and the potential $u_k$ become non-polynomial functions, which prohibits a perturbative RG analysis of this second order phase transition. We can rule out, however, that it falls into the directed percolation universality class since the characteristic rapidity inversion symmetry \cite{Janssen2005} is broken at any $k>0$ by a non-polynomial potential. It has been shown that deviations from a second order polynomial potential are RG relevant in $d=1$ \cite{NoiseFRGI} and thus will persist on the largest wavelengths, denying the restoration of rapidity inversion and the directed percolation universality at $k=0$.

\section{Persistence of the discontinuous dark state transition in higher dimensions}
\begin{figure}
\centering
\includegraphics[width=0.5\textwidth]{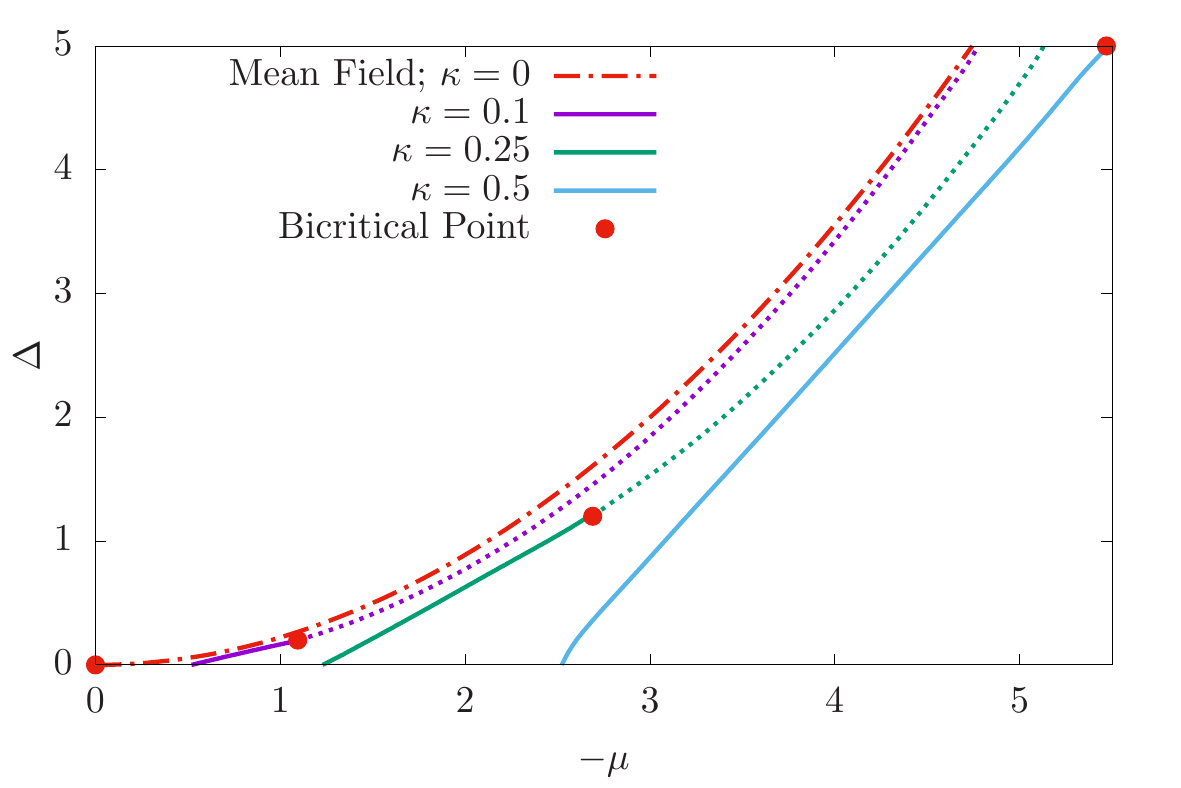}
\caption{Phase diagram of the three-dimensional system in the $\Delta-\mu$ plane for fixed $\lambda=1$. The mean-field first order transition (dot-dashed line) is gradually changed to second order (solid lines) upon increase of the initial noise level. First order phase transitions are generally present beyond mean-field (dotted lines), giving rise to a bicritical point at $\Delta \neq 0$. Increased initial noise levels also lead to a larger inactive phase. }
\label{3DPD}
\end{figure}

In higher dimensions, spatial fluctuations generally lead to less drastic modifications of the underlying mean-field picture. Only for very small potential barriers, the phase transition follows the same mechanism as in one dimension and becomes second order. We focus on intermediate to large potential wells, where we find that the first order transition between the dark and the finite density state persists in the presence of noise and spatial fluctuations for $d\ge2$ (see Fig.~\ref{3DPD}). For such initial potential barriers the double well structure of the potential persists up to the largest wavelengths and $V_{k}$ becomes convex only in the limit $k\rightarrow0$, establishing a single minimum. At the transition, the minimum jumps between $\nu_X=0$ and $\nu_X=\nu_f$ discontinuously such that the transition is of first order. This change in the potential evolution is accompanied by a drastic modification of the noise vertex flow compared to the case of a second order transition and the emergence of a sharp momentum scale for the onset of the evolution. These observations match with the phenomenology of first order phase transitions that are driven by nucleation and growth of droplets.

\subsection{Droplet phenomenology}
In order to discuss the results from the fRG approach, one needs a better understanding of the dynamics at a first order phase transition and adopt the corresponding picture to the present nonequilibrium setting. We briefly review the conventional phenomenology and adapt it to the present setting.

In higher spatial dimensions, local field fluctuations are suppressed and typically modify the mean-field picture only on the quantitative level, i.e. the dynamics in higher dimensions is dominated by small fluctuations around the saddle points of the action. The homogeneous saddle point solutions $\nu_X=0$ and $\nu_X=\nu_f$ in Eq.~\eqref{DetMF}, and corresponding small fluctuations, do, however, not interpolate between the dark state and the finite density phase and one has to go beyond a homogeneous approach in order to describe the discontinuous transition.

In the common picture \cite{BinderRev, Langer1}, first order phase transitions are driven by the nucleation and the growth of  droplets. Droplets are meta-stable field configurations which locally interpolate between the two different phases. To illustrate this, say the system is initialized in the finite density phase $\nu_{x,t=0}\approx \nu_f$ for all $x$. Tuning the system towards the transition by increasing the noise strength (or lowering the potential barrier) induces local transitions between the finite density solution $\nu_X=\nu_f$ and the zero density configuration $\nu_X=0$. Since spatial fluctuations are strongly suppressed, these transitions will only pass an inhomogeneous saddle-point 
\eq{DropSad}{
0\overset{!}{=}\frac{\delta S}{\delta\tilde{\nu}_X}=V'(\nu_X)-D\nabla^2\nu_X,
}
which interpolates between the finite density phase and a noise induced excitation to the top of the potential barrier $\nu_X=\nu_\text{max}$, see Fig.~\ref{figPot}. For a droplet centered around $x=x_0$ this is expressed via the boundary conditions
\eq{DropBound}{
\nu_X&=\nu_f \text{ for } |x|\rightarrow\infty,\\
\nu_X&=\nu_{\text{max}} \text{ for } x=x_0.
}
This configuration then deterministically reaches the dark state and nucleates a droplet.

Equation~\eqref{DropSad} requires local balance of kinetic and potential energy and the most relevant configurations are those which are most likely activated by the noise $\chi$. Since the present noise is field dependent, the structure of the most likely droplets will be a function of all the parameters, including the initial noise strengths. Their precise behavior cannot be determined quantitatively from a saddle point equation alone. Qualitatively, however, its solutions may still be expected to convey valuable information. What all of them have in common is that they are smooth and form a sharp domain wall separating the dark state configuration inside the droplet from the finite density region outside the droplet \cite{Coleman} (blue line in Fig.~\ref{figPot}). This is in contrast to the strongly fluctuating field configurations (red line in Fig.~\ref{figPot}) dominating the long-wavelength dynamics in one dimension. 

\begin{figure}
\centering
\includegraphics[width=0.5\textwidth]{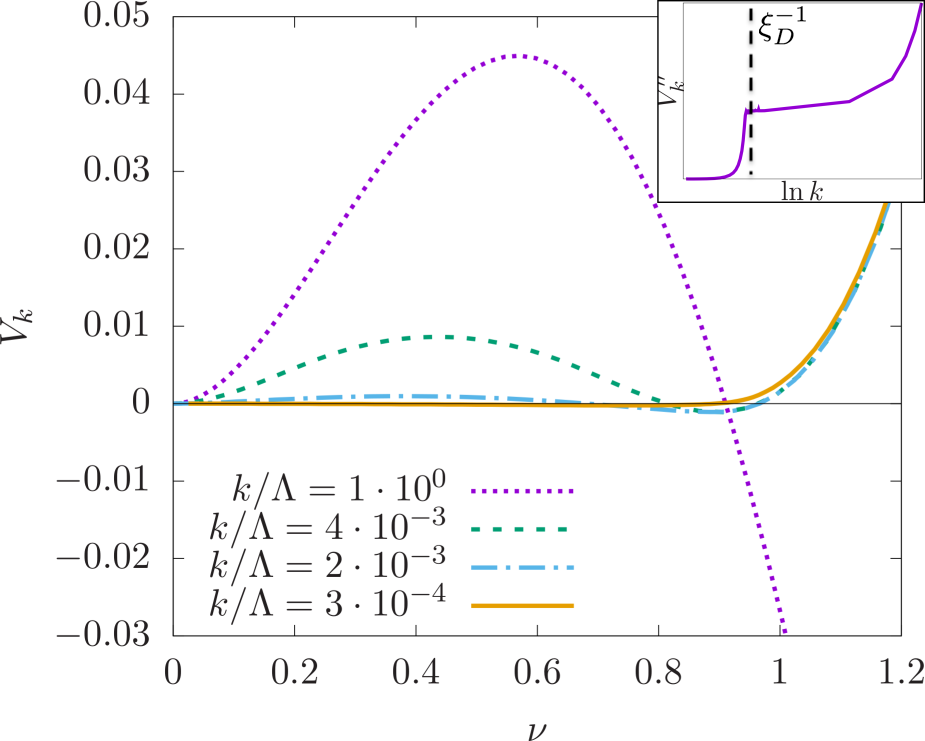}
\caption{Evolution of the deterministic potential $V_k$ for $\lambda = 1$, $\mu = -2.33$, $\Delta = 1$ and $\kappa = 0.1$. Convexity is achieved only after the fate of the nontrivial minimum is decided. The inset shows the RG evolution of $V_k''(0)$, providing a measure for the non-convexity. The droplet scale $\xi^{-1}_D$ is indicated by a sudden sharp drop.}
\label{TransPot}
\end{figure}

In this phenomenology, there exists a minimal extent of a droplet $\xi_D$, such that no droplet is formed below this size and thus no transitions between the minima occur on length scales $x<\xi_D$. While the value of $\xi_D$ depends in principle on all microscopic parameters and cannot be determined analytically, we observe the emergence of an extremely sharp, intermediate momentum scale $k_D$ during the RG flow. It marks the sudden onset of a fast evolution of both $V_k$ and $\chi_k$, see inset of Fig.~\ref{TransPot}. Associating this sudden onset with the formation of the smallest possible droplet allows us to identify $k_D\approx \xi_D^{-1}$ within the limits of our approach~\footnote{Note that this identification can only be made on a qualitative level, since the RG scale $k$ is regulator dependent and not an observable quantity.}. Appositely, this scale is absent at the second order transition. 

Once a droplet of the dark state phase is formed, its evolution underlies the initial Langevin equation \eqref{eq32}.
For a large, homogenous droplet in dimensions $d>2$, this equation is dominated by the potential energy $V(\nu_X)$. It will grow or shrink with a deterministic velocity $v_D=-\Delta V=V(\nu_f)$, i.e. collapse for $V(\nu_f)<0$ and expand for $V(\nu_f)>0$. This picture neglects spatial fluctuations at the domain walls and noise. Within the fRG approach, however, noise and spatial fluctuations are integrated out and lead to the continuous renormalization of $V_k$ and $V_{k\rightarrow0}(\nu_f)$ becomes the asymptotic droplet velocity.

We confirm this saddle-point picture of the transition in dimensions $d\ge2$ even in the presence of spatial fluctuations and noise and identify indeed a first order phase transition between the dark and the finite density phase for a large parameter regime. The transition is, however, in some aspects different from a first order transition at (thermal) equilibrium, which we will briefly discuss and address qualitatively in the following:
\begin{itemize}
\item The nucleation probability of droplets in a dark state region $\nu_X\approx0$ is strongly suppressed by the small noise level $\sim \nu_X$ and vanishes completely for the dark state. In the absence of additional noise channels, an initial dark state always remains dark.
\item The noise kernel prefers the dark state over the finite density state, i.e. there are much larger fluctuations inside than outside of a droplet. Noise and spatial fluctuations are therefore expected to modify $v_D$ and the phase boundary noticeably.
\item The saddle-point basin of attraction for the dark state encompasses densities $0\le \nu_X\le \nu_\text{max}$. The noise in this regime is, by construction, generally suppressed. One might ask whether the corresponding, small nucleation rate can induce significant transitions to the finite density state and whether the extended basin of attraction for the dark state will persist in the presence of fluctuations or not.
\end{itemize}

\subsection{Droplets within the fRG framework}
The clear evidence for a first order phase transition in the asymptotic fRG flow and the corresponding emergence of a sharp intermediate momentum scale for the onset of the flow strongly support the droplet interpretation of the first order transition. Still, it is not immediately obvious if and how the peculiarities of spatially inhomogeneous droplet solutions are accounted for within our RG framework, since we always project onto spatially constant density profiles. However, this does by no means inhibit fluctuations which are integrated over to acquire an arbitrarily complicated spatial structure. While we thus do not have a direct, quantitative handle on verifying the droplet picture, we will discuss further distinctive features of the RG flow that match very well with this framework.

\begin{figure}
\centering
\includegraphics[width=0.5\textwidth]{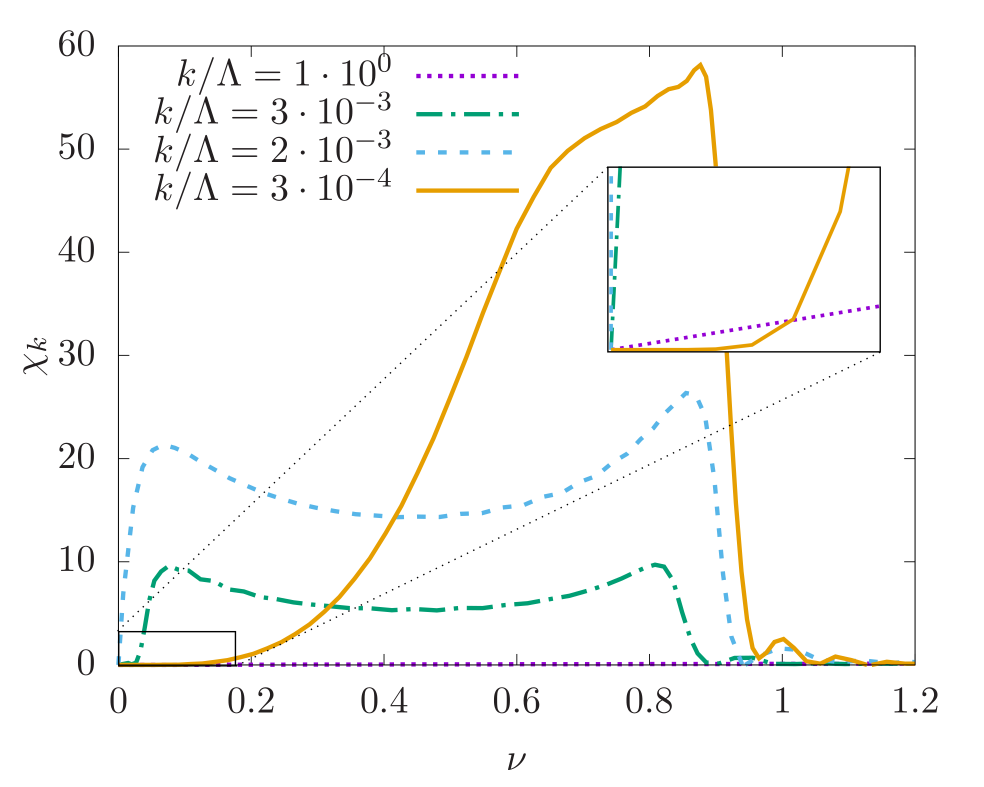}
\caption{Evolution of the noise kernel $\chi_k$ for $\lambda = 1$, $\mu = -2.33$, $\Delta = 1$ and $\kappa = 0.1$. Upon removal of the potential barrier, $\chi_k$ grows by three orders of magnitude and exhibits a bimodal structure. Only in the deep IR, fluctuations in the vicinity of the initial dark state are suppressed again. The inset provides an enhanced view on this event.}
\label{TransNoise}
\end{figure}

Let us first consider the flow of the deterministic potential for a configuration very close to a first order phase transition, see Fig.~\ref{TransPot}. While the potential shows significant evolution on intermediate and large scales, the potential barrier is removed only after the fate of the nontrivial minimum is decided~\cite{Note1}. 
This is in contrast to a second order transition, where long-wavelength fluctuations account for considerable corrections or even destruction of the nontrivial minimum even after the barrier is removed
\footnote{This enables us to uniquely identify a first order phase transition within our RG framework in the first place, as the discontinuity in the value of the order parameter is often not sufficiently distinctive to discriminate between a first and a steep second order transition.}.
This constitutes a strong hint on the nature of the fluctuations: as long as a potential barrier exists, droplets are well-defined excitations. In the vicinity of the first order transition, these droplets therefore exist until the fate of the nontrivial minimum is decided in an RG sense and the transition is droplet-driven.

However, these findings do not yet guarantee that these particular field configurations are in fact dominating the physics of the phase transition. In order to bring further clarification, we consider the flow of the noise kernel $\chi_k$ as well, see Fig.~\ref{TransNoise}. First we note that the noise amplitude grows by almost three orders of magnitude within a comparatively small window of RG scales~\cite{Note1}. This happens precisely at the scale $k_D\approx\xi_D^{-1}$, discussed in the previous section, where the potential barrier is in the process of being removed, cf. inset of Fig.~\ref{TransPot}. Second, as long as the flow has not reached the deep infrared, the noise kernel exhibits a very peculiar bimodal structure with the peaks roughly located at the positions of the nontrivial minima of the deterministic potential. This is exactly what one would expect for fluctuating droplet configurations, which suddenly jump from one minimum of the potential to the other, i.e. between the phases, and generate large field fluctuations around these minima. 

Only when the regime of extremely long wavelengths is reached, droplets begin to subside and fluctuations in the vicinity of the initial dark state become suppressed in order to facilitate the finite density phase~\footnote{While this particular feature is equivalent to what happens at a second order phase transition, it is here much more complicated to deal with numerically due to the steepness of the noise profile. This is the main reason why the expansion of the dark state in Fig.~\ref{TransNoise} seems to be less pronounced than in Fig.~\ref{1DNoiseEvo} where the flow could be run to much smaller scales.} as well as phase coexistence at the transition (see inset of Fig.~\ref{TransNoise}).

This interpretation of the noise kernel is further strengthened when reconsidering the one-dimensional case. As argued in sec.~\ref{sec1d} above, droplets are not expected to be the dominant fluctuations driving the phase transition. Indeed, the evolution of the noise kernel in Fig.~\ref{1DNoiseEvo} and also in the zero-density phase (not displayed here) does not display the bimodal behavior encountered in the vicinity of a first order phase transition.

\subsection{Coexistence point} 
Another aspect of the dark state transition is the formation of generically expected phase coexistence at the first order transition point, and the question whether a finite density on the one hand and the absence of fluctuations on the other can coexist. 

Following the fRG analysis one finds that exactly at the transition both the deterministic force and the noise kernel vanish for an extended set of field configurations $0\le\nu_X\le\nu_c$ (cf. Figs.~\ref{TransPot} and~\ref{TransNoise} inset). 
We now argue that this finding is generic to first order dark state phase transitions and neither an artifact of our theoretical approach nor a specific feature of this setup.
For any initial noise and potential configuration, the effective action $\Gamma_{k=0}$ must be a convex function of the fields $\nu_X, \tilde{\nu}_X$. For the dark state configuration both the noise kernel $\chi_k(\nu_X=0)=0$ and the potential $V_k(\nu_X=0)=0$ are, however, pinned to zero for all $k$. Requiring the coexistence of the dark and the finite density state at the transition, i.e. the absence of any deterministic or noise induced drift from one to the other, under the above conditions must therefore lead to a flat potential and vanishing noise kernel for an extended field configuration. 

\begin{figure}
\centering
\includegraphics[width=0.5\textwidth]{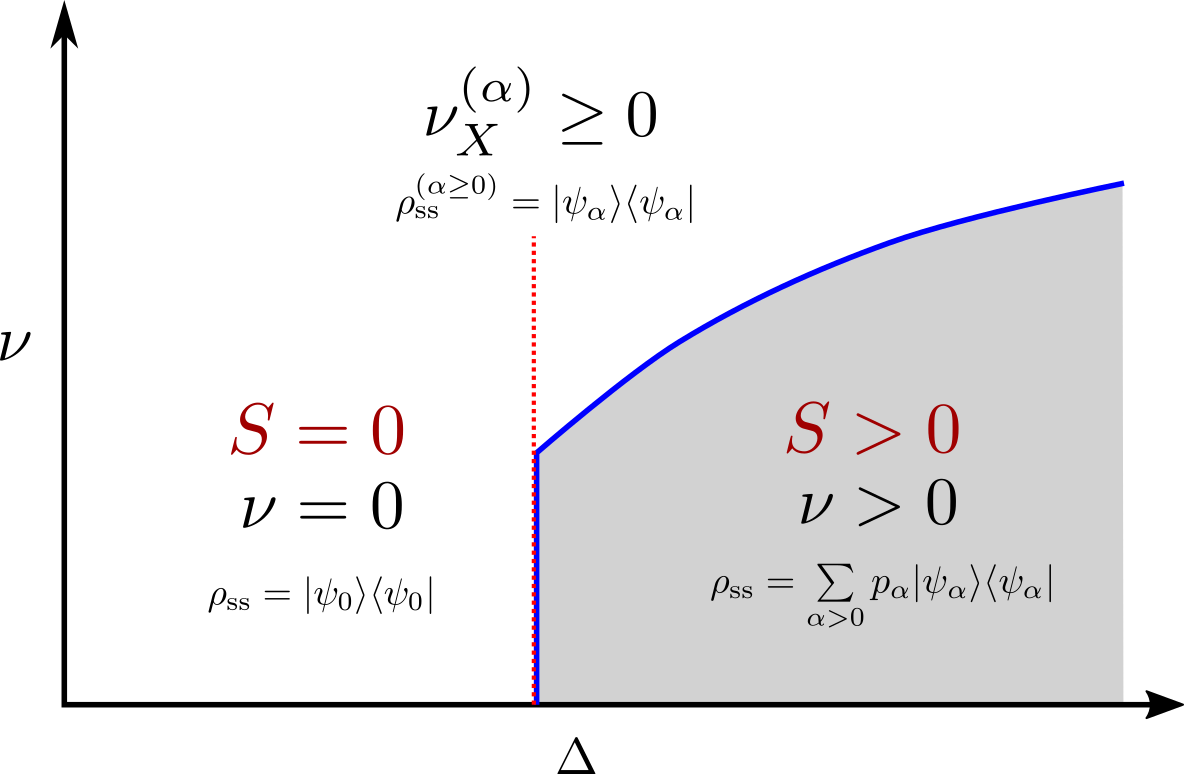}
\caption{Illustration of the first order dark state transition as a function of $\Delta$ in the field theory and in the Langevin equation picture. The dark state phase is characterized by a vanishing density field $\nu=0$ of spin-up atoms and a pure steady state $\rho_\text{SS}=|\psi_0\rangle\langle\psi_0|$, Eq.~\eqref{eq10}. In the finite density phase, the steady state has finite density $\nu>0$ as well as non-vanishing entropy $S$ and noise level.
Precisely at the transition, all field configurations $0\le\nu\le\nu_c$ represent noiseless steady state solutions, which we interpret as the presence of an extensive number of dark state configurations $\nu_X^{(\alpha)}$. Tentative analogies to fully quantum systems are indicated and discussed in the Outlook.}
\label{figDensMat}
\end{figure}

What we have to keep in mind, however, is that on the way from the master equation to the fRG results, the system has undergone a strong coarse graining procedure. A vanishing noise kernel in this coarse grained picture per se does not indicate that each individual spin is fluctuationless. Rather it shows the absence of fluctuations on thermodynamically large length scales. This specific feature of a first order dark state transition can be understood in the droplet picture. At the coexistence point, both phases are stable and one can thus insert a thermodynamically large, stable droplet, which will not expand, nor move, nor shrink considerably. The only expected fluctuations happen at the domain wall of the droplet, which is a sparse region. The absence of fluctuations may be interpreted such that away from the domain walls, the fields $\nu_X$ have exactly zero fluctuations, i.e. the atoms in this region are in an eigenstate of $n_l$. The domain walls instead may display finite density fluctuations but in the thermodynamic limit turn into a region of zero measure. Since the total number and size of droplets is arbitrary, one can construct arbitrarily many steady state field configurations $\partial_t\nu_X^{(\alpha)}=0$, that have a finite noise averaged density $\nu^{(\alpha)}=\frac{1}{V}\int_x \langle \nu_X^{(\alpha)}\rangle>0$ and at the same time vanishing fluctuations $\chi^{(\alpha)}=\frac{1}{V}\int_x \langle (\nu_X^{(\alpha)})^2\rangle-(\nu^{(\alpha)})^2=0$ in the thermodynamic limit of infinite system volume $V$, see Fig.~\ref{figDensMat} for an illustration.

We want to stress that for a flat potential and noise, the coarse grained Langevin equation becomes linear in $\nu_X$ and thus for any set $\{\nu_X^{(\alpha)}\}$ of solutions $\partial_t\nu_X^{(\alpha)}=0$, a linear combination $\nu^{(\vec{a})}_X\equiv \sum_\alpha a_{\alpha}\nu_X^{(\alpha)}$ with $a_{\alpha}\ge 0$ and $\sum_{\alpha}a_{\alpha}=1$ is a solution as well. The droplet configurations for different states $\alpha$ do not necessarily overlap in space and thus $\chi^{\vec{a}}\equiv\frac{1}{V}\int_x \langle (\nu_X^{(\vec{a})})^2\rangle-(\nu^{(\vec{a})})^2\ge0$ for average densities $\nu^{(\vec{a})}=\frac{1}{V}\int_x \langle \nu_X^{(\vec{a})}\rangle$. 

While a vanishing noise kernel and a vanishing potential at the coexistence point thus allow us to construct steady states with zero fluctuations, it does by no means enforce that the system relaxes towards such a state. Rather the $\nu_X^{(\alpha)}$ have to be understood as the basis set that spans the manifold of possible steady states. Depending on the specific choice of the $\{a_{\alpha}\}$, which is set by the initial conditions at $t=0$, these states interpolate between zero and finite fluctuations. The upper bound for the fluctuations is set by the noise kernel in the finite density phase. Connecting the strength of fluctuations with the system's entropy $S$, the coexistence of two phases with distinct entropy is thus explained with the coexistence of steady states $\nu_X^{(\vec{a})}$ with distinct fluctuation strength, ranging from strictly zero to a finite value.

This should be contrasted with a coexistence point at a thermal transition, where both the potential and the noise are as well flat  but the noise is proportional to the temperature $T$ of the system \cite{BinderRev, Langer1}. In principle, this allows us to perform the same construction as above for the steady state manifold but with fluctuations in the droplet states that are bounded from below by the temperature and thus never vanish. 

For any large but finite system with volume $V=L^d$ where $L$ is the linear dimension, the RG flow is cut off at momenta $k=L^{-1}$. While both $V$ as well as $\chi$ become flat for $k\rightarrow0$, for any $k\sim L^{-1}$ an asymptotically small potential well remains that separates the finite density from the zero density state. As a consequence, no deterministic path connects the two states and transitions between them can only correspond to noise activated trajectories. Since fluctuations on distances $x<L$ have all been integrated out, the only allowed noise activation trajectories are those on distances $x=L$, which have a rate $\gamma_A\sim \exp(-L^d\int_{\nu_f}^{\nu_\text{max}}\frac{V_k'}{\chi_k})$ \cite{Kamenev, NoiseFRGI}, that is exponentially suppressed in the volume and determined by the ratio of $V_{k}'/\chi_{k}$ at scale $k=L$. For finite systems, the degeneracy at the coexistence point is thus observably lifted on times $t>\gamma_A^{-1}$, which can for instance be associated with the finite system gap of the Lindbladian at the transition.

\section{Outlook}
We analyzed the dark state phase transition, i.e. the transition from statistically mixed to a pure steady state density matrix, in a spin model that is motivated by current experiments on driven dissipative Rydberg ensembles. After identifying a suitable order parameter for this transition, which underlies statistical fluctuations in the mixed state phase but is noiseless in the dark state, we demonstrated that the steady state dynamics of the order parameter can well be analyzed in terms of a nonequilibrium functional renormalization group approach. It allowed us to determine the phase boundary between the dark and the mixed state in the presence of noise and spatial fluctuations on a quantitative level, and to identify the nature of the transition (first or second order). Furthermore, we were able to identify the relevant fluctuations that drive the particular transition in the long-wavelength limit, and to establish the phenomenology for the dark state transition based on a droplet model and rare fluctuations.

This analysis sheds light on fluctuation induced dynamics close to dark state transitions and introduces a tool for the analysis of general first order phase transitions, particularly suited for non-thermal setups. From a methodological point, one might therefore ask whether the present functional renormalization group approach can be improved to enable, for instance, the analysis of systems driven by a non-Markovian (quantum) noise.

It also poses a set of new questions concerning quantum dark state transitions. The identification of a fluctuationless steady state field configuration $\nu_X$ with a dark state in the master equation framework was made concrete for the $\nu_X=0$ ferromagnetic ground state. In order to complete this analogy, one might investigate the coexistence point closer in order to explicitly identify the fluctuationless steady state field configurations $\nu_X^{(\alpha)}$ with another set of pure dark states $\rho_{\mathrm{SS}}^{(\alpha\ge 0)} = \ket{\psi_\alpha}\bra{\psi_\alpha}$, that is emergent at the transition, see Fig.~\ref{figDensMat}. One might then ask whether a thermodynamically large number of dark states can exist in a macroscopic many-body system.

The current analysis focusses mainly on the dynamics close to a particular steady state, but is unable to trace the kinetics close to a first order phase transition, which is in general expected to display universal behavior. This includes the question, in which way the steady state is reached in time and whether one can observe a transient dynamical bistability as predicted from the mean-field picture. Addressing these questions, in addition to the validation of the coexistence of many dark states at the first order phase transition, are possible tasks for further analysis of the model in terms of a Lindblad master equation approach and, even more directly, in terms of experiments with driven Rydberg ensembles.

\acknowledgments We thank B. Ladewig for fruitful discussions. S.D. and D.R. acknowledge support by the German Research Foundation (DFG) 
through the Institutional Strategy of the University of Cologne within the German Excellence Initiative (ZUK 81) and S.D. support by the European Research Council via ERC Grant Agreement n. 647434 (DOQS).
D.R. is supported, in part, by the NSERC of Canada. M.~B. acknowledges support from the Alexander von Humboldt foundation.

%
%


\bibliography{NoiseRef}

\end{document}